\documentclass[aps,physrev,preprint,groupedaddress]{revtex4-2}
\usepackage{graphicx}
\usepackage{amsmath, amsthm, amssymb}
\usepackage{bm}

\usepackage{physics}
\usepackage{hyperref} 
\begin{document}

%\preprint{}

\title{Eccentric Accretion Disks in Active Galactic Nuclei}


\author{Hongping Deng}
\email[]{hpdeng353@shao.ac.cn}
%\thanks{}
\affiliation{Shanghai Astronomical Observatory, Chinese Academy of Sciences, Nandan Rd 80th, Shanghai 200030, China}


\date{\today}

\begin{abstract}
We report that moderately eccentric flows around supermassive black holes (SMBHs), formed via either circumnuclear gas accretion or tidal disruption events, generate eccentricity cascades (from $>0.8$ to 0.2 outward), explaining multiwavelength emission and variability in active galactic nuclei (AGNs). The flows' non-axisymmetric temperature structure explains non-axisymmetric dust sublimation fronts, distinct broad emission-line components, and their radial motions. The innermost broad-line region (BLR) links to the SMBH vicinity through highly eccentric streams that produce soft X-rays at periapsis. General relativistic precession further compresses these flows, generating a hard X-ray continuum near the SMBH. Precession of the eccentric flow drives optical/X-ray variability, reproducing the observed X-ray power spectral density and occasional X-ray quasi-periodic eruptions. We thus propose eccentric accretion disks as a physical AGN model that unifies the elusive BLRs and X-ray corona. This model will enable detailed anatomy of AGNs and maximize their potential as cosmological standard candles.
\end{abstract}

%\keywords{}

\maketitle

%\section{Introduction}
\section{Introduction\label{intro}}
%\subsection{}
%\subsubsection{}
Active galactic nuclei (AGNs), among the universe's most luminous objects \cite{schmidt1963,seyfert1943}, radiate across the entire electromagnetic spectrum. They are powered by gravitational energy released from material feeding central supermassive black holes (SMBHs) \cite{lynden1969galactic}. However, current AGN theory remains largely phenomenological, relying on envisioned structures \cite{antonucci1993unified,netzer2015revisiting} beyond accretion disks \cite{shakura1973black} to produce broad optical emission lines \cite{amorim2024size} (broad-line regions, BLRs) and X-ray emission \cite{kara2025supermassiveaa} (the corona). It struggles to explain accretion disk sizes \cite{morgan2010quasar,fausnaugh2016space}, optical line diversity \cite{strateva2003double,rusakov2023broad,steinhardt2012sdss,Du2018monitoring}, BLR radial inflows/outflows \cite{pancoast2014model,du2016supermassive,zhou2019fast}, and, crucially, rapid X-ray \cite{gonzalez2012x,miniutti2019nine,risaliti2002ubiquitous} and optical variability (line width \cite{tohline1976variation,penston1984evolutionary,lamassa2015discovery,trakhtenbrot20191es,guo2025changing,yang2018discovery}, line profile \cite{schimoia2012short,gezari2007long,kollatschny2023outburst}, inflow/outflow switch \cite{williams2020space,chen2023broad}) on timescales significantly shorter than the disk accretion time \cite{Antonucci2013quasars,lawrence2018quasar,ricci2023changing}. As the observed population of changing-look AGNs continues to grow \cite{ricci2023changing,yang2018discovery,yang2025galaxies,guo2025changing}, we would expect AGNs with extreme variability \cite{guo2020high} to experience repetitive accretion bursts \cite{wang2025dormancy}; otherwise, it would be difficult to observe so many short-timescale state changes. This challenges the foundation of quasi-steady viscous accretion in AGNs.

This paper moves beyond the axisymmetric AGN model to explore the dynamics and variability of global eccentric flows around SMBHs, as motivated by recent studies. Double-peaked broad emission lines are likely produced in BLRs with appreciable eccentricity \cite{eracleous1995elliptical,strateva2003double}. Modern dynamical models of BLRs typically include radial flows \cite{pancoast2014method} on highly eccentric orbits, characterized by $\theta_e$, a measure of the eccentricity ($e=\sqrt{1-\sin^4\theta_e}$). Some of the best-studied AGNs contain flows (sometimes inflows) with eccentricity $e>0.8$ (equivalently $\theta_e<50^\circ$) \cite{pancoast2014model,williams2020space,amorim2020spatially}. Eccentric accretion disks are also expected when SMBHs accrete non-axisymmetric circumnuclear gas \cite{Hopkins2024,Hopkins2025,guo2024magnetized} or debris from tidal disruption events (TDEs)~\cite{Guillochon2014ps1,shiokawa2015general,ryu2023shocks}. However, the implications of eccentric flow—especially the non-axisymmetric temperature structure (see Appendix \ref{ap:a1}\footnote{Python scripts used for solving the vertical compression and uniformly precessing mode are available at the online public repository, \url{https://doi.org/10.5281/zenodo.15710795}})—for AGN emission and variability remain poorly understood. We carried out a series of hydrodynamic simulations, resolving extreme vertical compression with a novel Lagrangian scheme (see Appendix \ref{ap:b}), to investigate the evolution and variability of eccentric flows far from and close to SMBHs (Table \ref{tab:1}\footnote{Visualizations of the simulations and key output snapshots are available at the online public repository, \url{https://doi.org/10.5281/zenodo.15710795}}), taken together to form a unified AGN model.

\section{Simulations of eccentric disks}

Far from the SMBH, the problem is scale-free, so we measure radii in an arbitrary semi-major-axis unit $a_0$ and simulate disks spanning $a\sim 1$--$2\,a_0$ (Appendix \ref{ap:b}).
We performed simulations with an untwisted, uniform eccentricity of $e=0.2$ (A2E02-SF) or $e=0.5$ (A2E05-SF, A3E05-SF) to investigate the very outer regions and intermediate radii of eccentric AGN disks, respectively~\cite{Hopkins2024}.
In the high-eccentricity runs, gas can undergo rapid inward transport and reach radii where general-relativistic apsidal precession becomes important~\cite{tejeda2013accurate}. There, the relevant length scale is the gravitational radius, $R_g\equiv GM/c^2$. To emulate (and accelerate) the inspiral from larger radii, we restart the simulations from selected scale-free snapshots and ramp $R_g$ from 0 to $0.01a_0$ over $10\pi\,\Omega^{-1}(a_0)$ to study how eccentric flow is ultimately accreted onto the SMBH (Appendix \ref{ap:c}).

\begin{table*}[!ht]
\centering
\caption{Simulations of eccentric accretion disks. The typical output time interval is the local dynamical time at the disk inner edge, $\Omega^{-1}(a_0)$, while the A2E05-t1 and A2E05-t2 test simulations restart from the corresponding snapshots and deliver ten times more frequent outputs. The simulations employ up to 51 million Lagrangian fluid elements to capture the subtle parametric instability (see Appendix \ref{ap:b}).}
\label{tab:1}
\footnotesize
\begin{ruledtabular}
\begin{tabular}{cccccc}
       Run & Initial range [$a_0$] & Initial eccentricity & Run time [$\Omega^{-1}(a_0)$] & Resolution & Comments\cr
    A2UP-SF (LR/HR) & $1$--$2$ & 0.1--0.3 & 500 & 5M/42M & Scale-free\cr
    A2E02-SF & $1$--$2$ & 0.2 & 1000 & 5M & Scale-free\cr
    A2E05-SF & $1$--$2$ & 0.5 & 200 & 42M & Scale-free\cr
    A3E05-SF & $1$--$3$ & 0.5 & 200 & 51M & Scale-free\cr
    A2E05 & $1$--$2$ & A2E05-SF $t=100$ & 300 & 42M & $R_g=0.01a_0$\cr
    A3E05 & $1$--$3$ & A3E05-SF $t=200$ & 300 & 51M & $R_g=0.01a_0$\cr
    A2E05-t1 & $1$--$2$ & A2E05 $t=53$ & 20 & 42M & Dense output\cr
    A2E05-t2 & $1$--$2$ & A2E05 $t=150$ & 20 & 42M & Dense output\cr
\end{tabular}  
\end{ruledtabular}
\end{table*}

\begin{figure*}
\centering
  \includegraphics[width=0.92\textwidth]{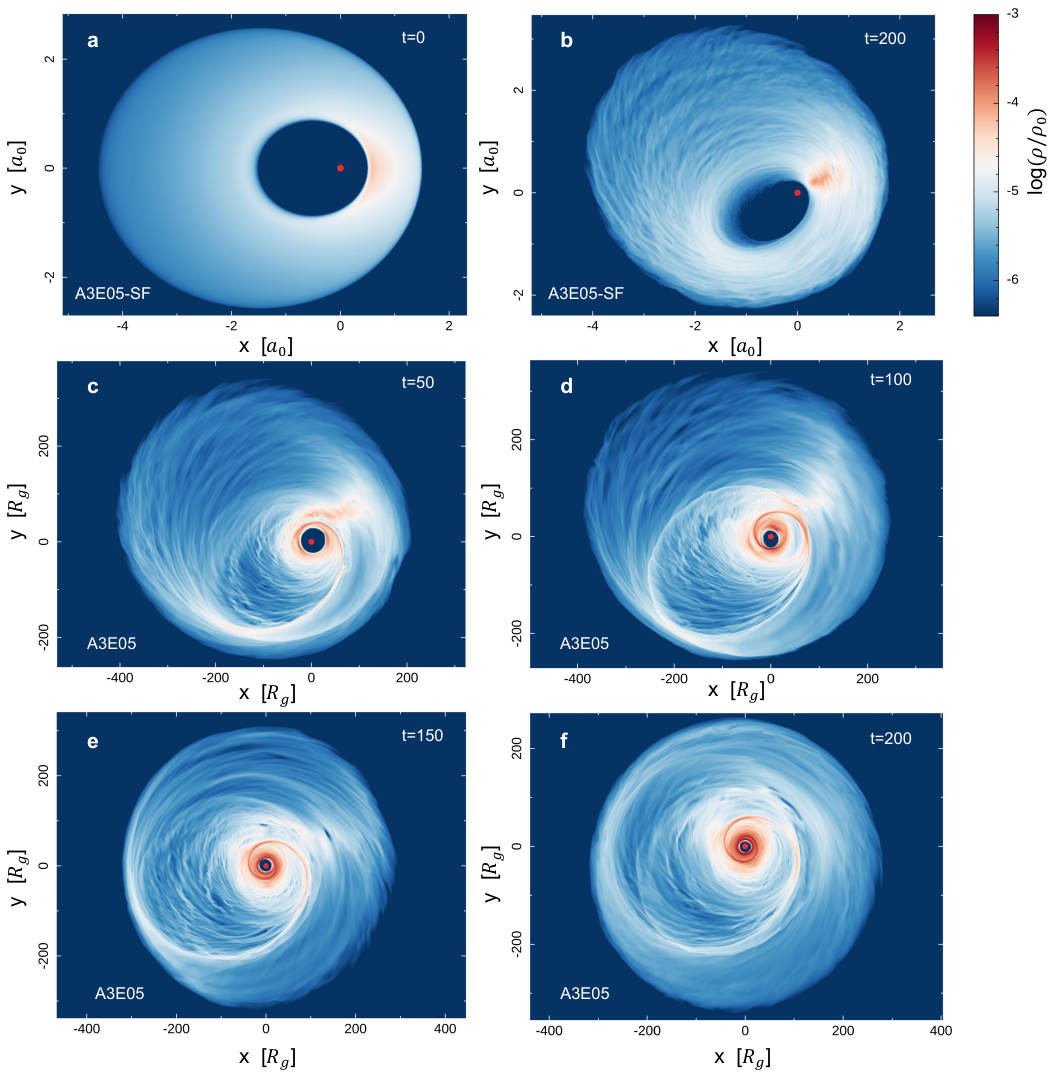}
  \caption{\textbf{Formation of highly eccentric flow and the resulting extreme compression.} Panels \textbf{a}--\textbf{f} show the midplane density (code units) for the scale-free simulation A3E05-SF and the A3E05 simulation with GR precession (Table \ref{tab:1}). Time is measured in units of the dynamical timescale at $a_0$, i.e., $\Omega^{-1}(a_0)$, where $a_0=100\,R_g$ for A3E05. The red circles denote the innermost stable circular orbit at $6\,R_g$. The highly eccentric flow undergoes strong compression and heating.}
  \label{fig:r3}
\end{figure*}
\clearpage

\begin{figure*}
  \centering
  \includegraphics[width=0.95\textwidth]{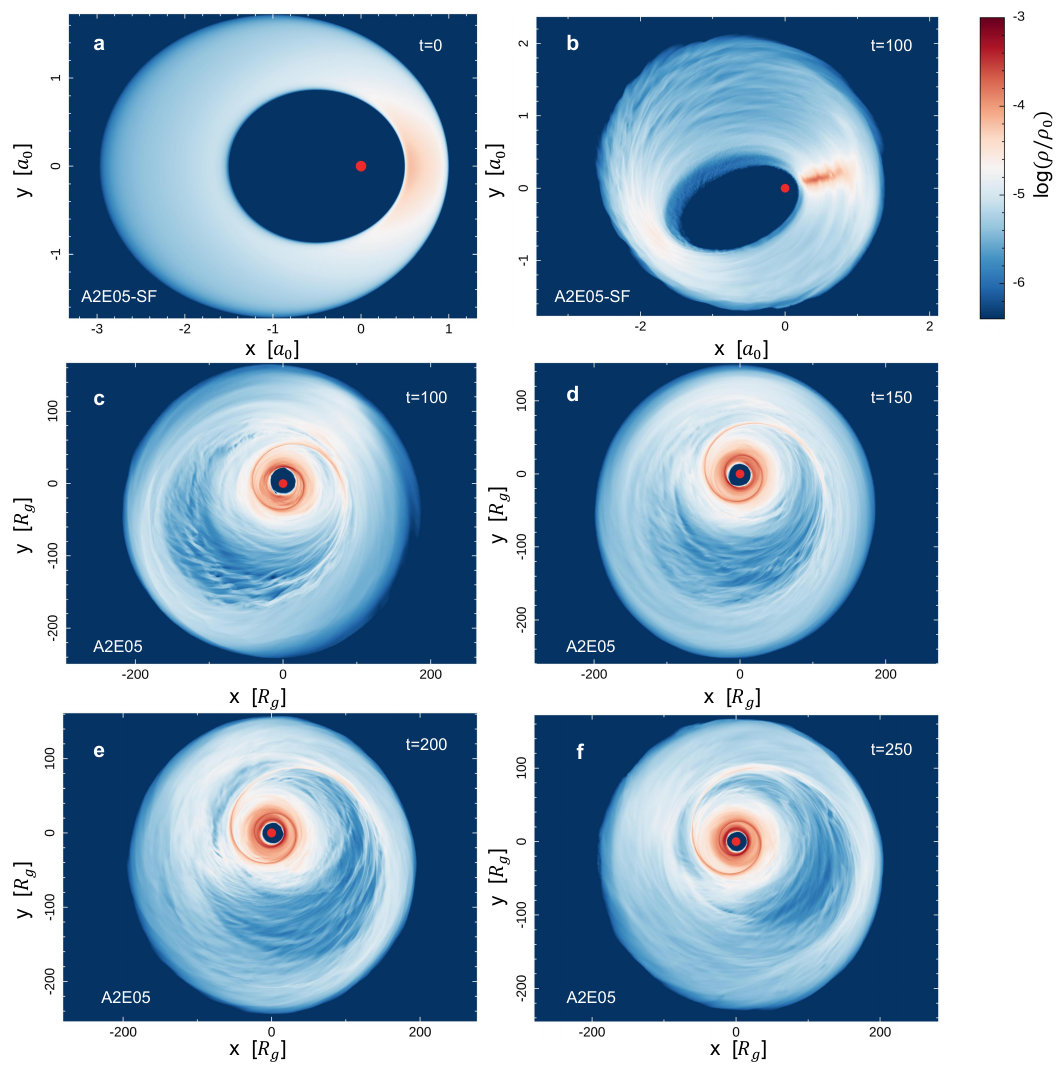}
  \caption{\textbf{Highly eccentric flow and extreme compression in a compact disk.} Panels \textbf{a}--\textbf{f} show midplane density maps analogous to Fig.~\ref{fig:r3}, but for A2E05-SF and A2E05 (Table \ref{tab:1}). In contrast to A3E05 in Fig.~\ref{fig:r3}, A2E05 reaches a quasi-global precessing state after $100\,\Omega^{-1}(100\,R_g)$.\label{fig:r2}}
\end{figure*}

\begin{figure*}
  \centering
  \includegraphics[width=0.9\textwidth]{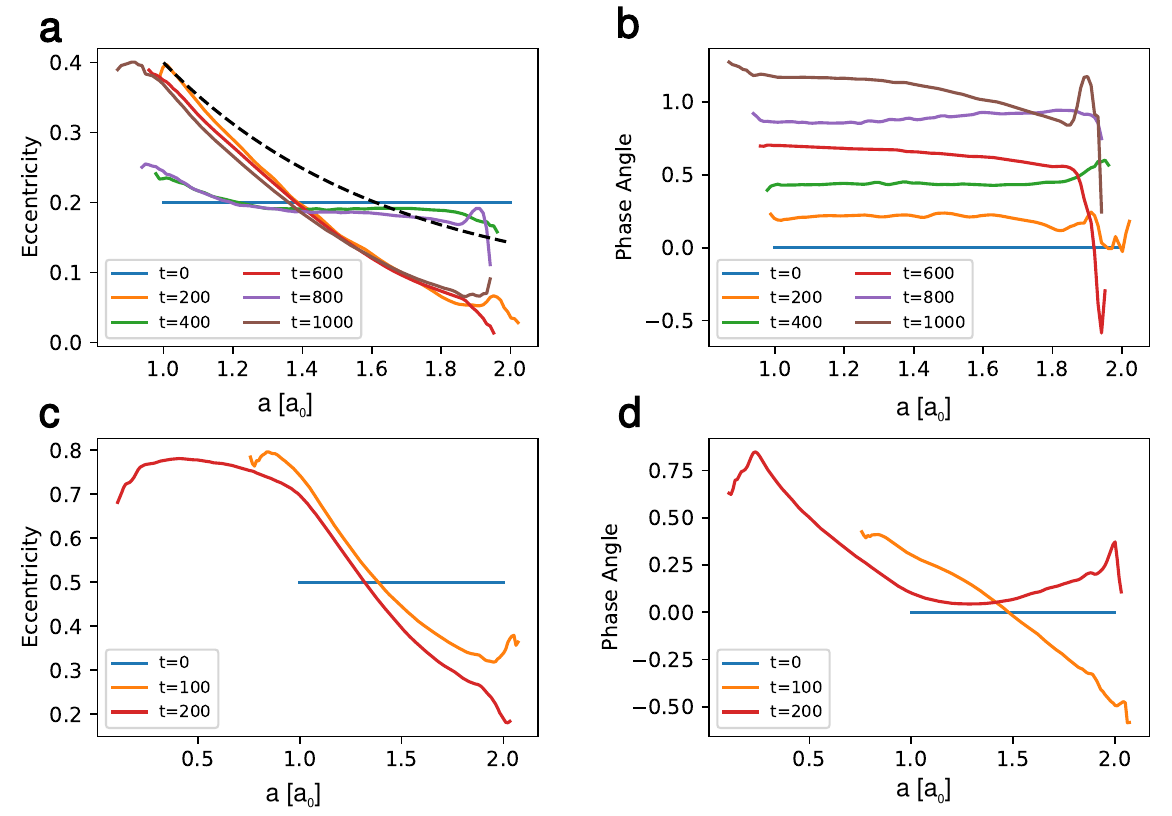}
  \caption{\textbf{Eccentricity cascade in low- and moderate-eccentricity disks.} Panels \textbf{a} and \textbf{b} show the evolution of the eccentricity profile and argument of periapsis for A2E02-SF. The eccentricity oscillates with minimal AMD damping every $200\,\Omega^{-1}(a_0)$ between a flat-eccentricity state and a state with a negative eccentricity gradient (the black dashed line in panel \textbf{a} shows a uniformly precessing model for reference; see Appendix \ref{ap:a2}). The twists are minimal in the two end-member states. However, the A2E05-SF model in panels \textbf{c} and \textbf{d} shows no such oscillations; instead, it exhibits rapid accretion, as indicated by the evolving material distribution.\label{fig:eprofile-sf}}
\end{figure*}

In the fiducial scale-free models (A3E05, A2E05), we consider a disk of moderate eccentricity ($e=0.5$), as typically found in AGN formation \cite{Hopkins2024,Hopkins2025} and TDE simulations~\cite{shiokawa2015general}, and find that it quickly develops a negative eccentricity gradient, reaching $e\sim 0.8$ at its inner edge (Figs.~\ref{fig:r3}a,b and \ref{fig:r2}a,b; Fig.~\ref{fig:eprofile-sf}). This occurs naturally because disks with decreasing eccentricity are less susceptible to orbital collisions and thus more stable than other configurations with the same angular momentum deficit (AMD) budget (see Appendices \ref{ap:a1} and \ref{ap:d}). Nevertheless, orbital collisions are unavoidable in the highly eccentric inner disk, leading to dissipation and coherent accretion (Fig.~\ref{fig:eprofile-sf}).

A similar eccentricity cascade is established in the low-eccentricity flow (A2E02-SF). However, orbital collisions are rare, and the flow oscillates between the initial flat-eccentricity state and a decreasing-eccentricity state with negligible mass accretion (Fig.~\ref{fig:eprofile-sf}). The observed eccentricity cascade (A3E05-SF, A2E05-SF) and eccentricity oscillation (A2E02-SF) can occur at essentially any radius far from the central SMBH, highlighting the restless nature of eccentric flows.

\begin{figure*}
  \centering
  \includegraphics[width=\textwidth]{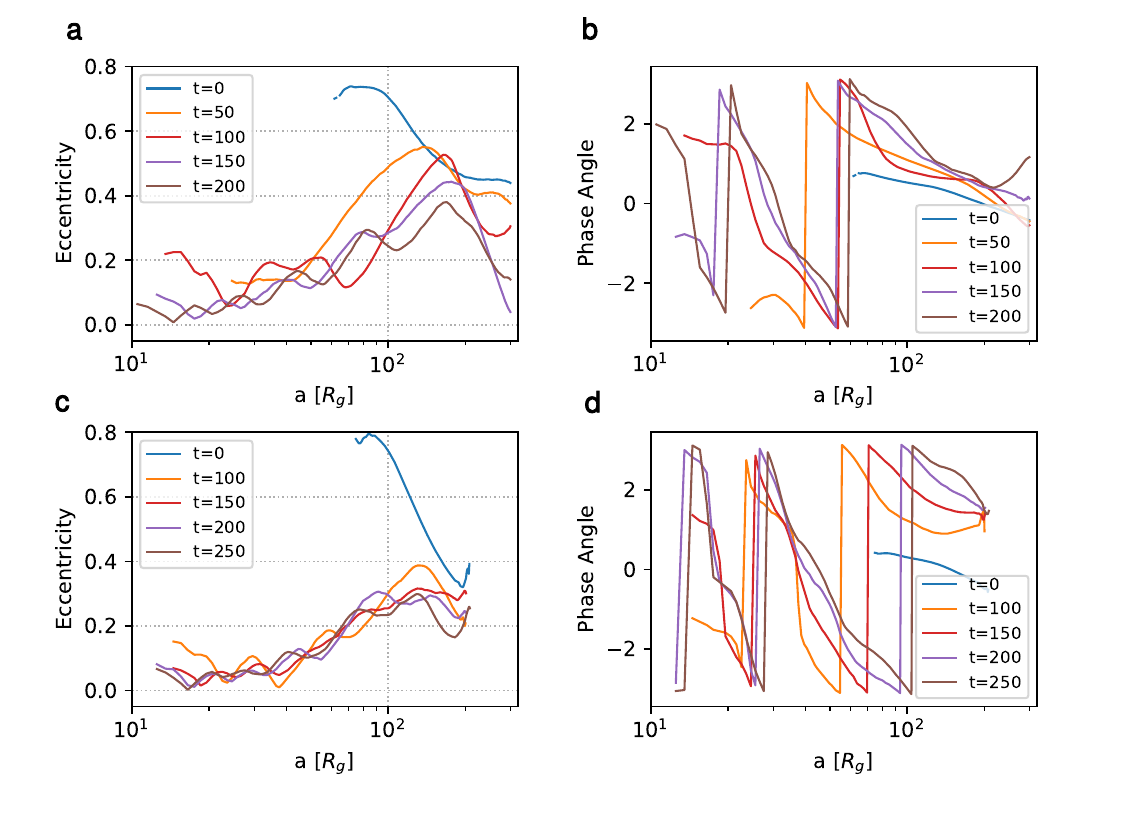}
  \caption{\textbf{Eccentricity profiles for A3E05 and A2E05.} Panels \textbf{a} and \textbf{b} show the eccentricity and argument of periapsis of the flow in A3E05 (see also Fig.~\ref{fig:r3}). We note that A3E05 starts from A3E05-SF (Table \ref{tab:1}; Fig.~\ref{fig:r3}b). Panels \textbf{c} and \textbf{d} show the corresponding quantities for A2E05, which starts from A2E05-SF (Fig.~\ref{fig:r2}b).\label{fig:eprofile}}
\end{figure*}

The coherent accretion observed in Fig.~\ref{fig:eprofile-sf}c\footnote{See an animation of A2E05-SF in the online Zenodo repository.} can bring material quickly to the vicinity of the black hole ($\sim 100\,R_g$) along highly eccentric orbits, where general relativistic (GR) precession becomes non-negligible. After turning on GR precession, A3E05 and A2E05 form non-closed orbits and spiral structures close to the SMBH (Figs.~\ref{fig:r3}c--f and \ref{fig:r2}c--f). This leads to further compression, increasing the density by more than an order of magnitude. The eccentricity of the innermost flow decreases toward the black hole and becomes nearly circular near $10\,R_g$, as shown in Fig.~\ref{fig:eprofile}. Notably, the eccentric disks are twisted in both the scale-free simulations and the simulations with GR precession; the latter show significant twists with complex evolution due to strong differential precession (Fig.~\ref{fig:eprofile}).

\begin{figure*}
  \centering
  \includegraphics[width=\textwidth]{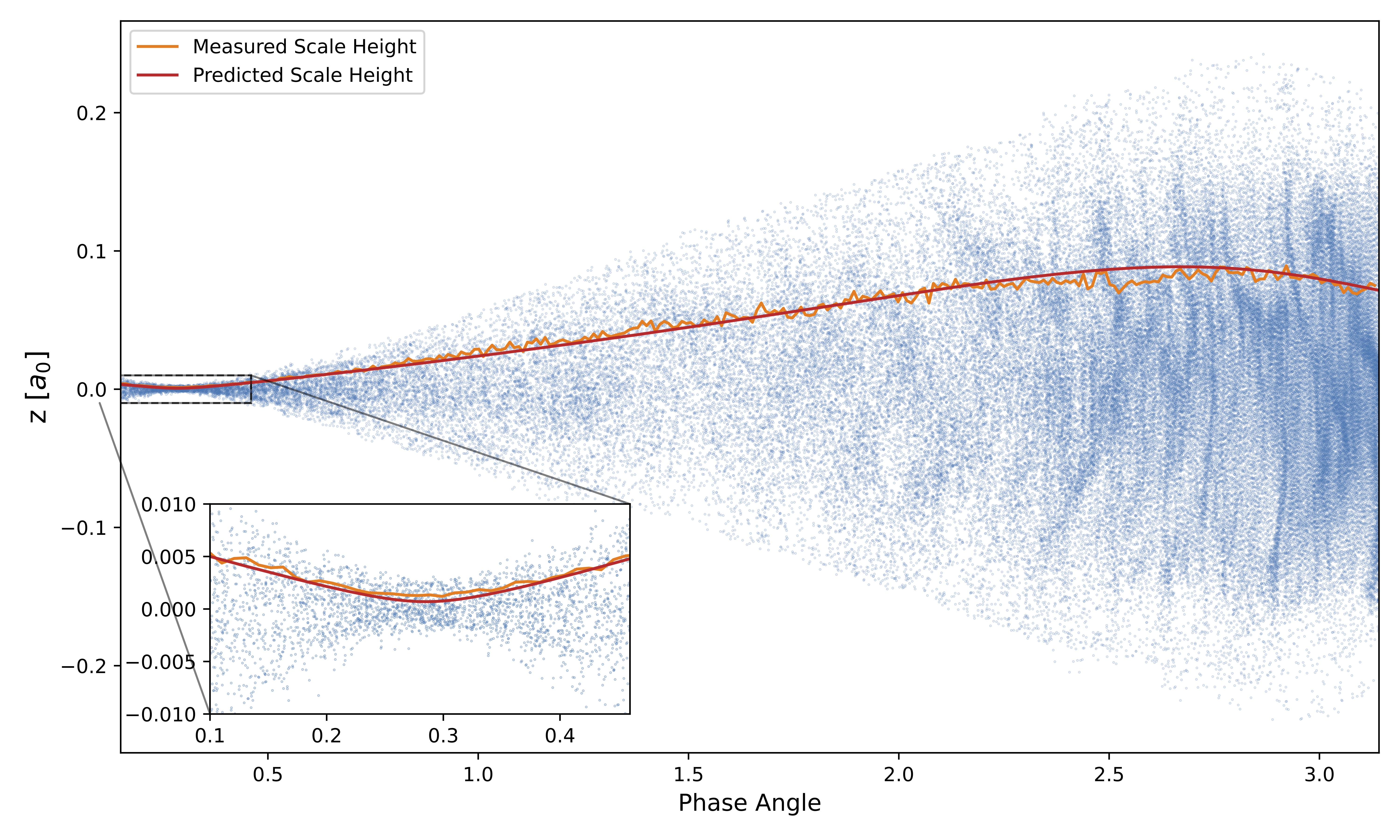}
  \caption{\textbf{Extreme vertical compression in eccentric flows.} The plot shows the distribution of Lagrangian fluid elements in $[1.18a_0,1.185a_0]$ for the eccentric flow in Fig.~\ref{fig:r2}b, where the measured local eccentricity, dimensionless eccentricity gradient, and twist are $(e, ae_a, a\omega_a)=(0.62,-0.76,-0.64)$. The disk is compressed by a factor of 127 near periapsis relative to apoapsis (true anomaly is shown in the plot), as predicted by eccentric-disk theory (see Appendix \ref{ap:a1}) and accurately captured by our Lagrangian method. Even more extreme compression is expected in inner AGN accretion disks, where high-eccentricity flow is common.\label{fig:H}}
\end{figure*}

The above simulations reveal key features of eccentric flows in AGNs far from and near the central SMBH. Taken together, we expect the eccentricity to peak at a semimajor axis of $\sim 500\,R_g$, reaching values $>0.8$, where GR precession becomes effective and eventually circularizes the flow interior to the most eccentric orbit (see Figs.~\ref{fig:r3} and \ref{fig:r2}). This is evident in the numerical experiments A3E05 and A2E05, although the limited dynamical range and AMD budget hinder the sustenance of highly eccentric flows (Fig.~\ref{fig:eprofile}). True AGN accretion disks~\cite{Hopkins2025} likely span $\gtrsim 10^5\,R_g$ (even post-TDE disks~\cite{shiokawa2015general,ryu2023shocks} cover thousands of $R_g$), providing enough space for an extended eccentricity cascade with large AMD reservoirs.

Unfortunately, it is currently unfeasible to cover such a large dynamical range while simultaneously resolving the disk vertical structure (e.g., Fig.~\ref{fig:H}\footnote{Similar vertical compression and expansion is reported in TDE simulations \cite{Guillochon2014ps1,ryu2023shocks,shiokawa2015general}, but is likely under-resolved.}), which critically affects eccentric-disk evolution (see Appendix \ref{ap:a2}). Instead, we use observational constraints to construct a reference eccentric-disk AGN model informed by the simulations (Table \ref{tab:1}) and the semianalytical eccentric-disk theory in Appendix \ref{ap:a}.

\begin{figure*}

  \centering
  \includegraphics[width=\textwidth]{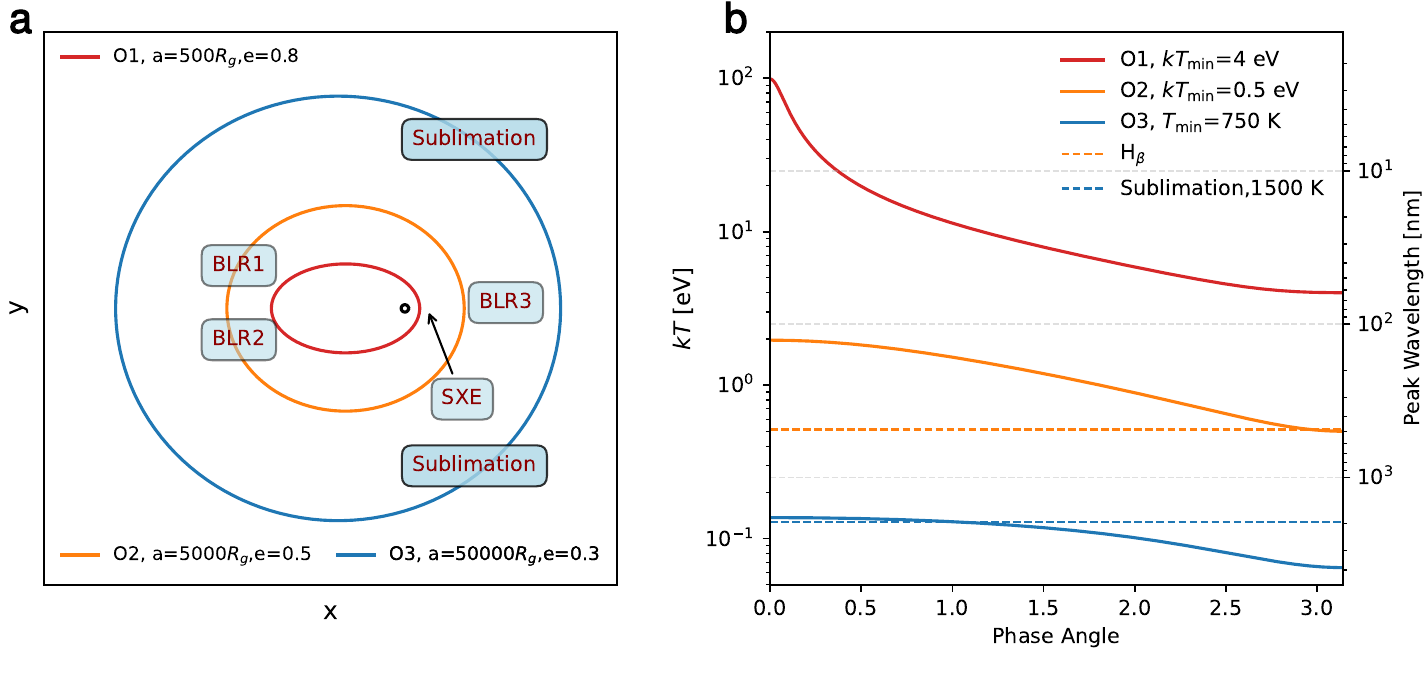}
  \caption{\textbf{Reference structure of AGN accretion disks.} From our simulations, which form highly eccentric flows but cover a limited dynamical range (Fig.~\ref{fig:r3}), we construct a reference eccentric-disk AGN model informed by observations of AGN physical scales and temperatures. \textbf{a.} Three reference orbits, O1, O2, and O3, with eccentricities 0.8, 0.5, and 0.3, and semimajor axes $500\,R_g$, $5000\,R_g$, and $50{,}000\,R_g$ (not shown to scale). \textbf{b.} Temperature variation along these reference orbits versus true anomaly, and the peak emission wavelength of the corresponding blackbody radiation. Orbit O1 connects the inner broad-line region (BLR) to the vicinity of the black hole ($100\,R_g$), producing soft X-ray excess (SXE) near periapsis. Orbit O2 produces broad H$\beta$ emission in two BLR patches near apoapsis, while exterior to O2 the H$\beta$-emitting temperature can be reached in BLR3. Along orbit O3, dust sublimates at $\sim 1500\,\mathrm{K}$ near periapsis, leaving asymmetric dust lanes.}
  \label{fig:rorbits}
\end{figure*}
\section{Unification of AGN structures}

The sizes of key emission regions in AGNs are commonly constrained by reverberation mapping, which measures time lags in response to luminosity variations of the central engine \cite{cackett2021reverberation}. In particular, the broad-line regions (BLRs) typically span $10^3\,R_g$--$10^5\,R_g$, lying between the conventional \emph{accretion disk} and the dusty torus \cite{netzer2015revisiting}, and are among the defining observational features of AGNs \cite{seyfert1943,cackett2021reverberation,pancoast2014model,du2016supermassive,williams2020space,amorim2020spatially}.

To connect our scale-free simulations to these observationally inferred scales, we construct a reference eccentric-disk model built around three representative BLR orbits (Fig.~\ref{fig:rorbits}): O1, O2, and O3 with semimajor axes $500\,R_g$, $5000\,R_g$, and $50{,}000\,R_g$, and eccentricities 0.8, 0.5, and 0.3, respectively.

\begin{figure*}
  \centering
  \includegraphics[width=\textwidth]{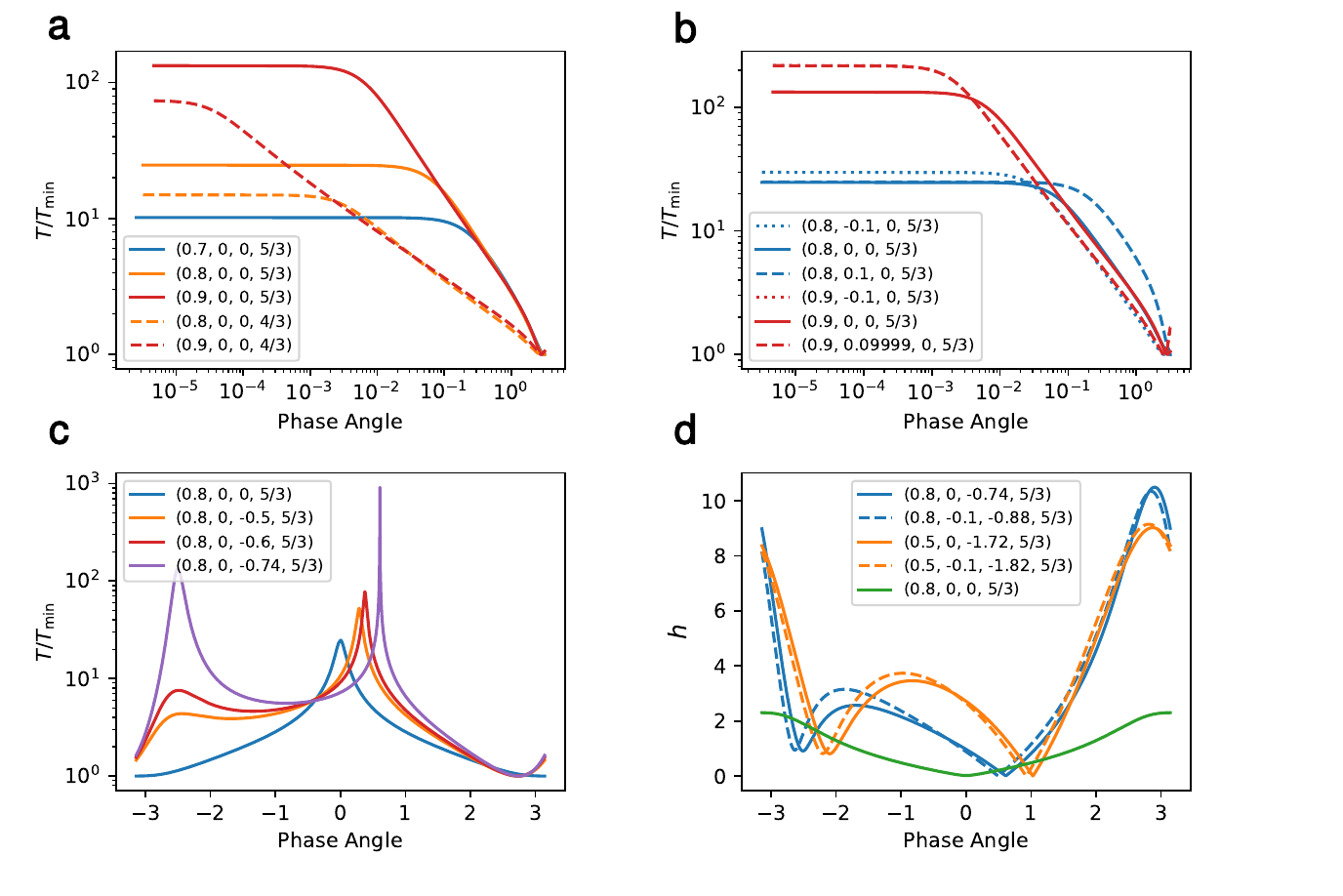}
  \caption{\textbf{Azimuthal temperature and disk thickness variations.} Eccentric disks are characterized by the eccentricity, dimensionless eccentricity gradient and twist, and the adiabatic index $(e, ae_a, a\omega_a, \gamma)$, as labeled in the legends. Panels \textbf{a} and \textbf{b} show disks with no twist ($\omega_a=0$), for which the temperature has reflection symmetry about the major axis. Panel \textbf{c} shows two peaks in the temperature profile for some strongly twisted disks. In general, eccentricity gradients or twists lead to stronger temperature variations than the aligned, constant-eccentricity disks considered in Fig.~\ref{fig:rorbits}. Near the critical configuration for orbital crossing (e.g., $(0.9, 0.9999, 0, 5/3)$ and $(0.8, 0, -0.74, 5/3)$), catastrophic vertical compression can occur (see panel \textbf{d} for scale-height factors), and the simple analytic solutions for the vertical structure admit no upper limit on the temperature (see Appendix \ref{ap:a1}).\label{fig:vare}}
\end{figure*}

Guided by the simulation trends (Fig.~\ref{fig:eprofile-sf}), we consider an interpretive scenario in which the outer disk near O3 episodically supplies high-AMD material to the O2 region ($e\sim 0.5$) during eccentricity oscillations, sustaining a more eccentric flow ($e\sim 0.8$) around O1 via an ``eccentricity cascade'' (Fig.~\ref{fig:eprofile-sf}c). Gas near O1, analogous to the most eccentric flow in Fig.~\ref{fig:r3}c, undergoes differential GR precession and can populate the vicinity of the black hole with dense, hot material. The quoted scales and eccentricities are intended as reference values and may vary with an AGN's formation history and evolutionary stage (Fig. ~\ref{fig:vare}). The flow temperature is regulated by central irradiation, accretion and compressional heating, and radiative cooling; because radiation reprocessing is beyond the scope of this paper, we adopt representative temperatures guided by observational ensembles.

\subsection{Hot dust emission in infrared}\label{sec:dust}
The reference orbit O3 (similar to Fig.~\ref{fig:eprofile-sf}a) marks the dust sublimation front, interior to which the temperature is too high for dust formation. We assume a minimum temperature $T_\mathrm{min}=750\,\mathrm{K}$ (at apoapsis) along orbit O3. Despite its low eccentricity, the temperature rises to more than twice this value near periapsis. As a result, dust near periapsis sublimates, leaving an incomplete elliptical dust lane, as observed in recent optical/near-infrared interferometric imaging \cite{pfuhl2020image,gamez2022thermal}. Flows exterior to orbit O3 are dusty and can be highly chaotic, feeding the accretion disk with high-AMD material \cite{Hopkins2025,guo2024magnetized}.

\subsection{Broad emission lines}
The intermediate reference orbit O2 (similar to Fig.~\ref{fig:r3}a) represents the bulk of the BLR. While eccentric disks have long been invoked to explain broad emission lines \cite{eracleous1995elliptical,strateva2003double}, many models adopt a prescribed emissivity that depends only on radius.\footnote{Non-axisymmetry then enters only indirectly, e.g., through stronger emission near periapsis than apoapsis.} In contrast, the vertical compression/expansion in eccentric disks generically produces strong azimuthal temperature variations (Fig.~\ref{fig:H}; Appendix \ref{ap:a1}) that should directly shape where and how efficiently lines are emitted.

As an illustration, we focus on the H$\beta$ line \cite{du2016supermassive,pancoast2014model,bentz2013low,williams2020space,chen2023broad,trakhtenbrot20191es,guo2025changing} and adopt a representative minimum temperature $kT_\mathrm{min}=0.5\,\mathrm{eV}$ (with $k$ the Boltzmann constant). Along O2, the temperature varies by a factor of $\sim 4$, selecting two apoapsis patches (BLR1 and BLR2; $kT_\mathrm{min}=0.51\,\mathrm{eV}$) where the local continuum peaks near the H$\beta$ wavelength and hydrogen is readily excited for subsequent de-excitation and line emission\footnote{Variations in irradiation can change the local energy balance and cause line-strength changes similar to continuum reverberation mapping. In fact, photoionization may dominate atom excitation at the disk atmosphere, given the vertical expansion at apoapsis (Fig.~\ref{fig:H}).}.

Exterior to O2, compressional heating near periapsis can again bring gas into the H$\beta$-emitting temperature range, defining a third emitting zone (BLR3). The $5100\,\text{\AA}$ continuum emission is primarily produced near orbit O2 \cite{jiang2024evidence}, and its luminosity should scale with the emitting area and hence approximately with the square of the BLR distance, consistent with observations \cite{bentz2013low,chen2023broad}.

More generally, the three emitting zones (BLR1--BLR3; Fig.~\ref{fig:rorbits}) offer a physical motivation for the three broad Gaussian subcomponents often used when fitting broad-line profiles \cite{strateva2003double,storchi2017double}. For an untwisted eccentric flow between O1 and O3, the centroid velocities of the two apoapsis components (BLR1 and BLR2, located at true anomalies $\theta_1$ and $\theta_2$ symmetric about the semimajor axis) and the periapsis component (BLR3) are
\begin{equation*}
 v_{1,2}=-v_0\frac{\sin \theta_{1,2}}{\sqrt{1-e^2}}\mathbf{\hat{e}_x}+v_0\frac{e+\cos\theta_{1,2}}{\sqrt{1-e^2}}\mathbf{\hat{e}_y},\quad v_3=v_0\sqrt{\frac{1+e}{1-e}}\mathbf{\hat{e}_y},
\end{equation*}
where $v_0=\sqrt{GM/a}$ and $\mathbf{\hat{e}_x}$ and $\mathbf{\hat{e}_y}$ are unit vectors along the major and minor axes. For a given line, the temperature structure set by the central radiation field determines where the emitting patches form, thereby selecting $v_0,e, \theta_{1,2}$ and fixing the relative Doppler shifts.

In this framework, the line profile is primarily controlled by $M$, $a$, $e$, $\theta_{1,2}$, and the viewing geometry, whereas obscuration modulates the relative strengths of the components. The radial motion of eccentric streamlines naturally produces apparent inflow/outflow signatures when viewed at different inclinations \cite{du2016supermassive,zhou2019fast}. Asymmetric profiles \cite{strateva2003double,storchi2017double,Du2018monitoring} and large peak velocity offsets \cite{rusakov2023broad} can likewise arise, and may be further amplified if particular BLR components are preferentially obscured.

\begin{figure*}
  \centering
  \includegraphics[width=\textwidth]{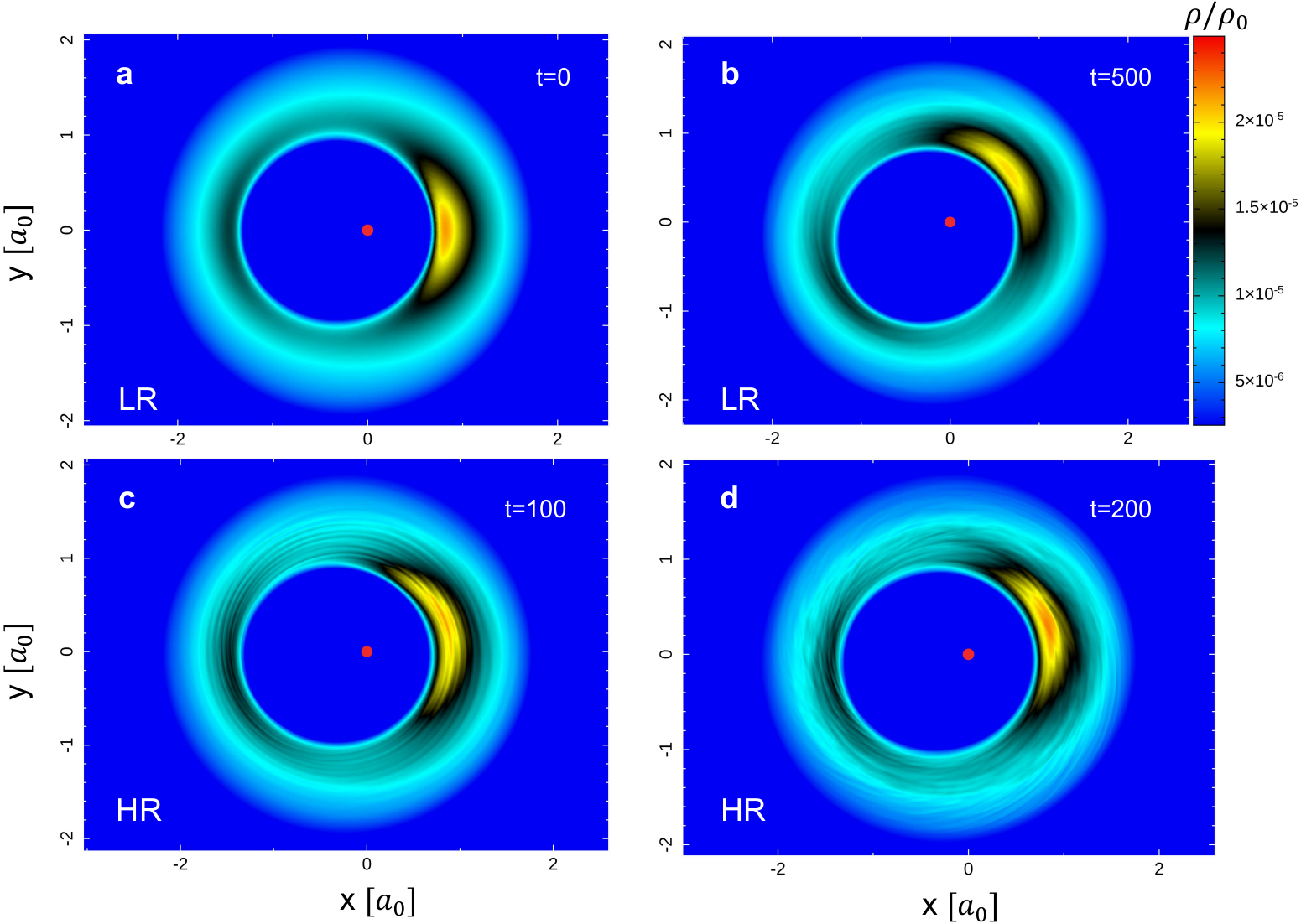}
  \caption{\textbf{Hydrodynamic instability in eccentric disks}. In panels \textbf{a} and \textbf{b}, the midplane density in the low-resolution A2UP-SF model shows no sign of turbulence. The high-resolution realization of A2UP-SF develops prominent corrugation patterns (panel \textbf{c}) by $100\,\Omega^{-1}(a_0)$ and vigorous turbulence by $200\,\Omega^{-1}(a_0)$ (panel \textbf{d}).\label{fig:upress}}
\end{figure*}

\begin{figure*}
  \centering
  \includegraphics[width=0.7\textwidth]{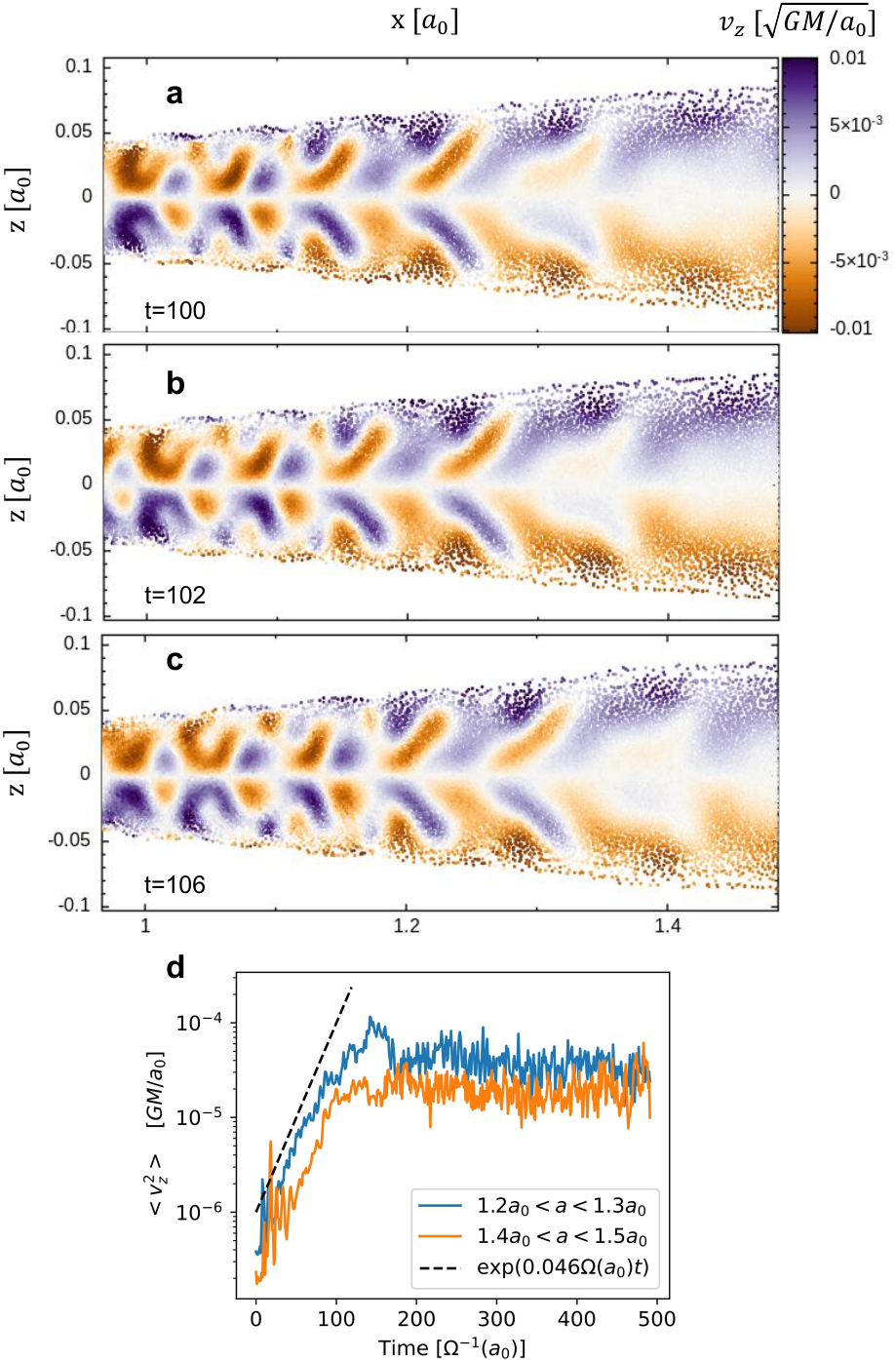}
  \caption{\textbf{The development of turbulence}. Panels \textbf{a}--\textbf{c} show the vertical flow velocity in a thin slice of thickness $0.01\,a_0$ along the periapsis axis for A2UP-SF (HR), which should be zero in the absence of instability. Around $100\,\Omega^{-1}(a_0)$, wave packets traveling at an angular frequency $\sim\Omega (a_0)$ become prominent, indicating parametric instability. Panel \textbf{d} shows the growth of the instability, quantified by the average of $v_z^2$ in such slices at different distances. \label{fig:pi}}
\end{figure*}

\begin{figure*}
  \centering
  \includegraphics[width=\textwidth]{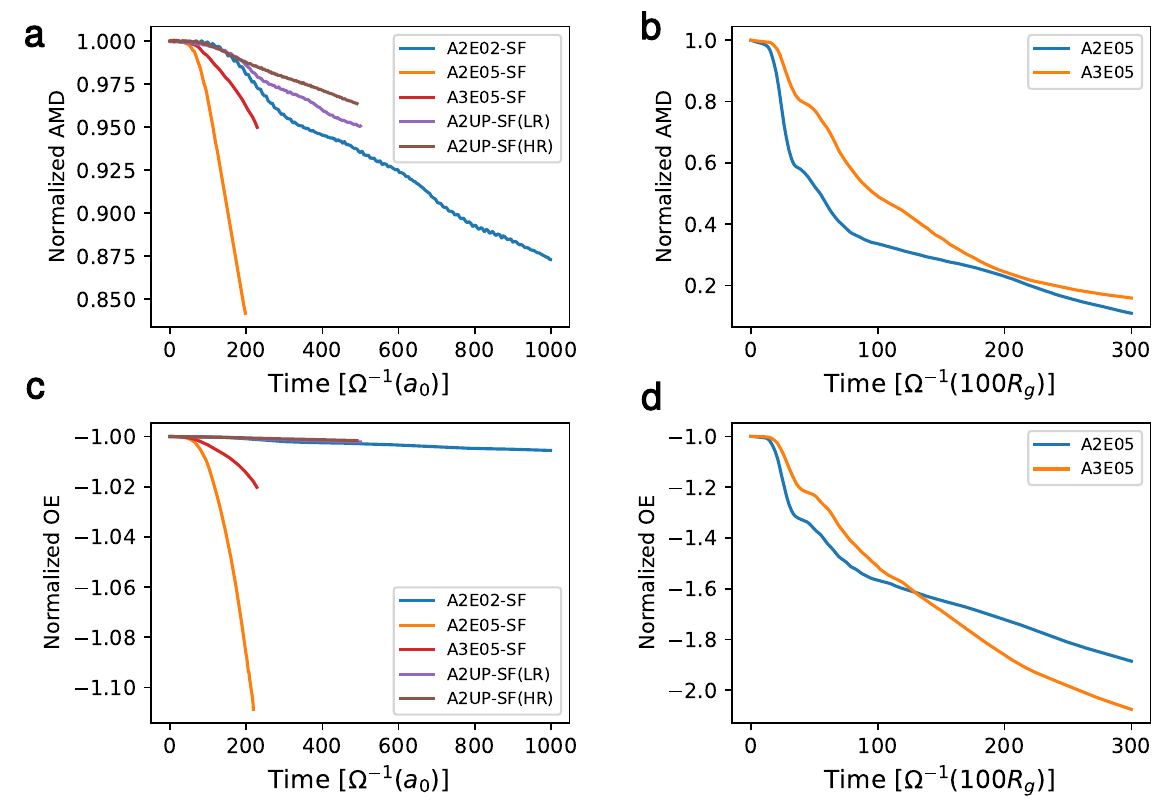}
  \caption{\textbf{AMD and accretion}. Panels \textbf{a}--\textbf{d} show the fractional changes in the angular momentum deficit (AMD) and orbital energy (OE). The parametric instability (Figs.~\ref{fig:upress} and \ref{fig:pi}) leads only to slow accretion, whereas collisional shocks in A2E05-SF and A3E05-SF facilitate rapid accretion (Fig.~\ref{fig:eprofile-sf}). GR precession in A2E05 and A3E05 drives restless precession of highly eccentric flow and repeated orbital collisions, resulting in strong accretion. The AMD budget is used up quickly in our simulations because of the limited dynamical range and the lack of sustained supply, in contrast to extended real AGN disks (Fig.~\ref{fig:rorbits}).\label{fig:amd}}
\end{figure*}

\begin{figure*}
  \centering
  \includegraphics[width=\textwidth]{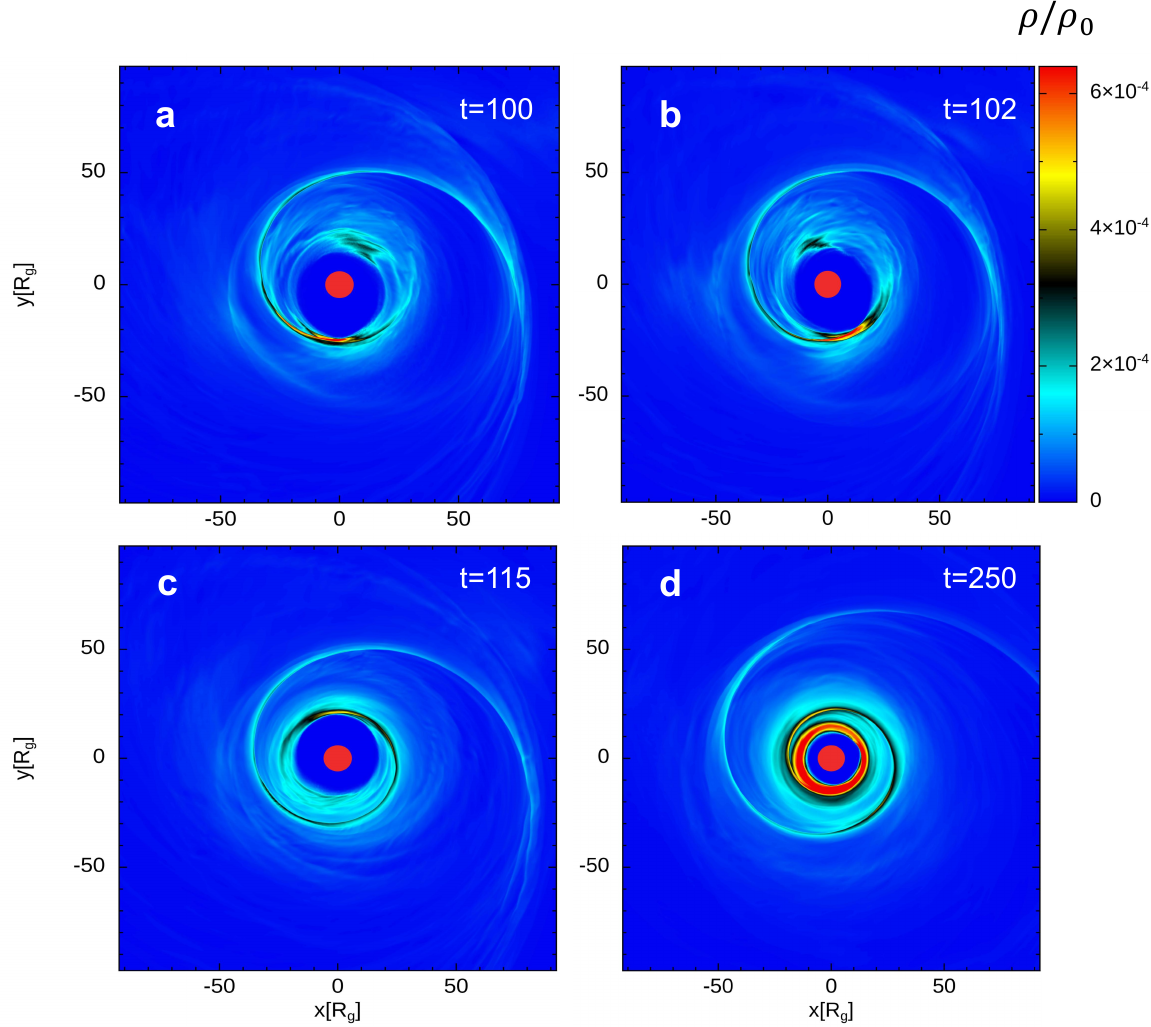}
  \caption{\textbf{Destruction and reformation of the densest X-ray core in A3E05}. Panels \textbf{a}--\textbf{d} zoom in on the midplane gas density around the black hole, sharing a linear color scale. They highlight the extra compression of the flow after periapsis passage ($\sim 100\,R_g$) along the reference orbit O1 (Fig.~\ref{fig:rorbits}), driven by GR precession. In Fig.~\ref{fig:r3}d, the most eccentric flow collides with the outer, less eccentric flow at $t=100\,\Omega^{-1}(100\,R_g)$; after about one full period of the most eccentric flow, the core is strongly disrupted at $t=115\,\Omega^{-1}(100\,R_g)$. The core gradually strengthens as accretion continues, although we lack a proper treatment of mass accretion within $10\,R_g$.\label{fig:core}}
\end{figure*}

This framework also suggests qualitative trends in broad emission lines. At high eccentricity, BLR3 is expected to be subdominant because its emitting area is small (Appendix \ref{ap:a1}) and because the periapsis region exhibits a steep temperature gradient (Fig.~\ref{fig:rorbits}b). This provides a natural route to the lopsided BLR geometries inferred in several well-characterized AGNs \cite{pancoast2014model}. In addition, the photocenters of BLR1 and BLR2 should be separable with modern optical interferometry and offset relative to the hot-dust photocenter, which is itself non-axisymmetric (see section \ref{sec:dust}) \cite{amorim2024size}.

At low eccentricity (and correspondingly weak X-ray emission; Fig.~\ref{fig:vare}a), the line-of-sight velocities across the BLR are reduced, favoring narrow observed line widths and, in the extreme, a population of true Type~II AGNs lacking hidden broad lines (non-HBLR S2s) \cite{tran2003unified,merloni2014incidence,netzer2015revisiting}. Orientation further modulates these phenomena~\cite{shen2014diversity}: at relatively low inclinations (near face-on), the projected velocities are minimized, and rapidly accreting objects (e.g., driven by highly eccentric flow) may be classified as narrow-line Seyfert~1s and even non-HBLR S2s \cite{zhang2006unified}. %misalignments between disks and radio jets inferred observationally \cite{runnoe2013orientation} support a range of viewing geometries.

When the AGN luminosity decreases, the BLR contracts (smaller semimajor axis), increasing the characteristic line width \cite{wang2020sloan}. The relative contribution from BLR3 is then further reduced, while the superposition of BLR1 and BLR2 can produce a large projected velocity separation along $\mathbf{\hat{e}_x}$. If the observer lies within the corresponding (small) range of viewing angles, prominent double-peaked profiles \cite{eracleous1995elliptical,strateva2003double,schimoia2012short,wang2005two} should be detected.

\subsection{UV and X-rays}

The reference orbit O1 (similar to the most eccentric orbit in Fig.~\ref{fig:r3}e) marks the outer edge of the conventional \emph{accretion disk}, which generates the UV/optical continuum \cite{netzer2015revisiting,cackett2021reverberation}. In contrast to the circular accretion disk model, additional UV emission can be produced efficiently near the periapsis of orbit O2, explaining the UV excess or the so-called big blue bump \cite{malkan1982ultraviolet,vanden2001composite,lawrence2012uv}.
Due to the high eccentricity, the \emph{accretion disk} size at a given temperature is about twice that of a circular disk \cite{fausnaugh2016space,morgan2010quasar,cackett2021reverberation}.

The extreme vertical compression (Fig.~\ref{fig:H}) near the periapsis of O1 can easily reach a high enough temperature to generate soft X-ray blackbody emission (Fig.~\ref{fig:rorbits} and \ref{fig:vare}). This provides a natural explanation for the almost universal soft X-ray excess \cite{bianchi2009caixa} in AGN, with a blackbody temperature of about 0.1 keV. Higher eccentricity leads to a more significant soft X-ray excess, as well as more frequent orbital collisions (see Appendix \ref{ap:a}) and thus faster accretion (higher Eddington ratio) \cite{boissay2016hard}.

Gas compressional heated near $\sim 100R_g$ is then transported inward and further compressed along the GR-precession-driven spiral structure (Fig.~\ref{fig:r3} and Fig.~\ref{fig:r2}), generating the hard X-ray continuum (2--10~keV) within $\sim 20R_g$. Because the UV- and X-ray-emitting gas are connected by the same family of eccentric orbits inside O1 (Fig.~\ref{fig:r3}), their luminosities are expected to be tightly correlated \cite{lusso2016tight}. In this sense, the hydrodynamic eccentric disk model (for the effects of radiation and magnetic fields, see Appendix \ref{ap:e}) can produce the observed X-rays without invoking an additional, ad hoc corona \cite{netzer2015revisiting,kara2025supermassiveaa}.

\section{AGN variability}
\subsection{Optical variability}

The static eccentric-disk model above provides useful scales and structures for interpreting multi-wavelength AGN emission. Eccentric disks, however, are intrinsically time-dependent. They experience differential precession driven by internal pressure forces and by external torques, including GR (Appendix \ref{ap:a2}; Figs.\ \ref{fig:eprofile-sf} and \ref{fig:eprofile}). They are also unstable to parametric instabilities when pairs of inertial waves resonate with the orbital motion \cite{barker2014hydrodynamic}. In an ideal disk that precesses uniformly (Fig. \ref{fig:upress}), we find exponential growth of this instability, which drives hydrodynamic turbulence (Fig. \ref{fig:pi}) but with an effective viscosity $\alpha< 0.001$ (Appendix \ref{ap:g}). Similar turbulence develops in A305-SF (Fig.~\ref{fig:r3}b) and A205-SF (Fig. \ref{fig:r2}b).

Pressure-induced precession operates on a timescale shorter than viscous transport by a factor $\sim 1/\alpha$ (Appendix \ref{ap:a}). The resulting differential precession promotes orbital crossings, shocks, and rapid inward transport (Fig. \ref{fig:amd}). Nevertheless, in our simulations the outer, low-eccentricity flow near orbits O2 and O3 is circularized (Fig. \ref{fig:amd}a) and accreted (Fig. \ref{fig:amd}c) only after thousands of local orbits, comparable to the lifetime of AGN flickering \cite{schawinski2015active} and to the growth phase, respectively.

These instabilities therefore cannot account for the observed \emph{fast} optical variability \cite{schimoia2012short,lamassa2015discovery,chen2023broad,guo2025changing}. We instead propose that GR-driven precession of the flow near orbit O1 is the key driver. For an eccentric orbit, the GR apsidal-precession rate is
\begin{equation*}
    \dot{\omega}_{\rm GR}=\frac{3R_g n}{a(1-e^2)},
\end{equation*}
where $n=v_0/a$ is the mean motion (Appendix \ref{ap:c}). The O1 orbits precess by one radian in $1.05\,(M/10^7M_\odot)$ years. The associated precession of the inner flow---often manifesting as global spiral structure \cite{storchi2017double} (Figs.~\ref{fig:r3}c--f and \ref{fig:r2}c--f)---naturally explains the observed year-scale variability of double-peaked broad lines \cite{schimoia2012short,eracleous2009double,gezari2007long}.

Precession of the most eccentric orbit can also affect broad emission lines produced further away when its apoapsis collides with the BLR flow represented by orbit O2\footnote{In the reference model presented in Fig.~\ref{fig:rorbits}, there is no direct collision with orbit O2 after orbit O1 precession. However, true AGN possess a continuous eccentricity cascade and may well have an O1--O2 orbit collision when the BLR is compact for relatively low-luminosity AGNs.} (Fig.~\ref{fig:r3}cd). The collision shocks will first lead to an increase in optical/UV emission, and the perturbation will propagate inward, leading to the temporary disruption of the region of hard X-ray continuum emission (Fig.~\ref{fig:core}). Meanwhile, the outer flow increases in eccentricity (Fig.~\ref{fig:eprofile}a), forming broader emission lines and undergoing a potential transit from a type II to type I AGN if the line width surpasses the somewhat arbitrary threshold \cite{netzer2015revisiting} of 1000 km/s. The more eccentric outer flow also leads to a stronger UV excess at periapsis, explaining the general bluer-when-brighter trend in AGN optical variability \cite{ruan2014evidence,yang2018discovery}. Similar turn-on processes are observed in the changing-state AGN 1ES 1927+654 \cite{li2022host,trakhtenbrot20191es,ricci2020destruction}. We note that whether the outer flow loses or gains angular momentum depends sensitively on the velocity at the contact point, which is determined by the eccentricity cascade. When the outer flow gains angular momentum, it decreases in eccentricity, causing a type I to type II AGN transition \cite{lamassa2015discovery,ricci2023changing,guo2025changing}.
\begin{figure*}
  \centering
  \includegraphics[width=0.8\textwidth]{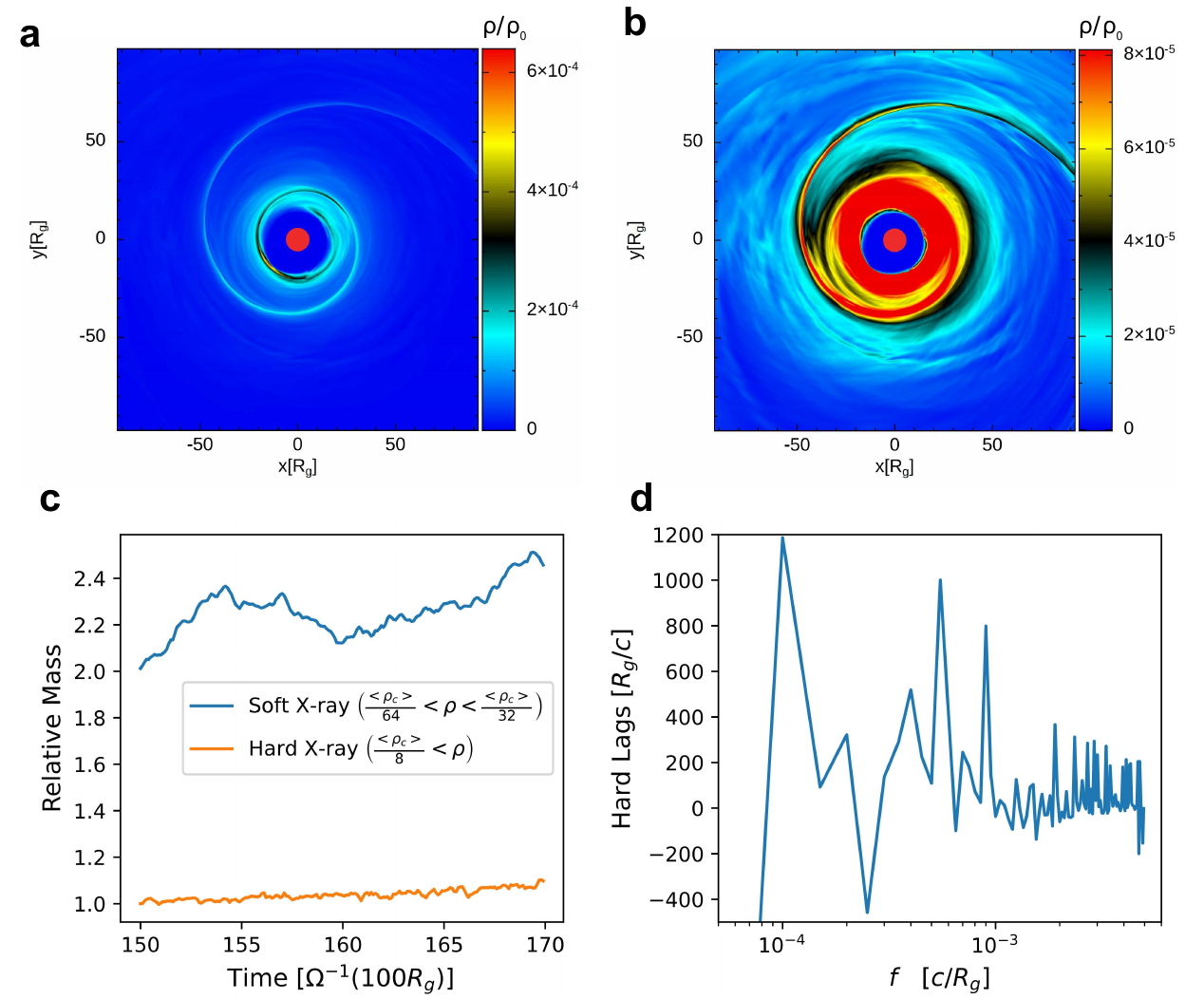}
  \caption{\textbf{Hard and soft X-ray emission}. Panels \textbf{a} and \textbf{b} highlight the central region of A2E05 at $t=150\,\Omega^{-1}(100\,R_g)$ (Fig.~\ref{fig:r2}d), similar to Fig.~\ref{fig:core}. Here, we use two color scales to highlight the soft X-ray excess (cyan regions in panel \textbf{b}) contributed by material near the periapsis of  the most eccentric flow in Fig.~\ref{fig:r2}d. This excess is evident in panel \textbf{c} for A2E05-t2, where the soft X-ray-emitting region is twice as massive as the hard X-ray region and covers a much larger area, despite its narrow density range ($\langle\rho_c\rangle$ is the time average of $\rho_c$ over the interval shown). The hard X-ray variation (estimated from the mass variation in panel \textbf{c}; see Appendix \ref{ap:f}) lags behind the soft X-ray variation over the frequency range explored.\label{fig:hardsoft}}
\end{figure*}

\begin{figure*}
  \centering
  \includegraphics[width=\textwidth]{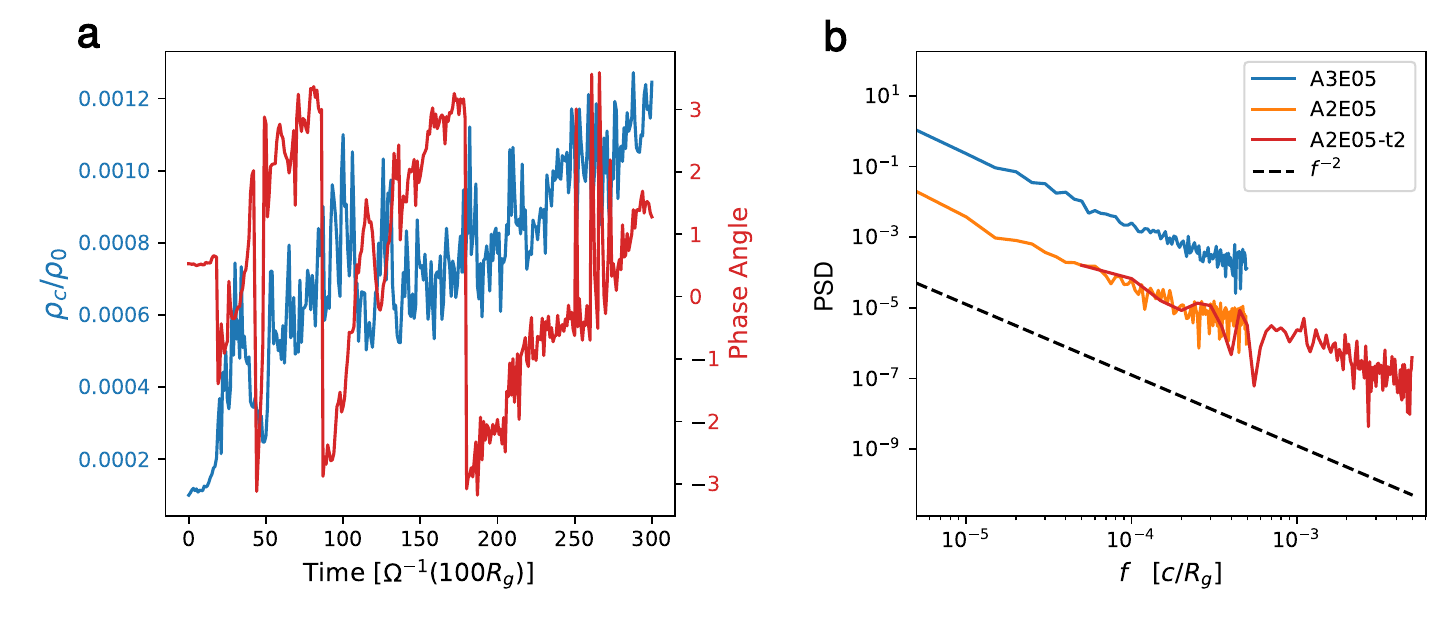}
  \caption{\textbf{Variability of X-ray emission.}
  \textbf{a} The average density and phase angle of the core in A3E05 (see also Fig.~\ref{fig:qpe}). \textbf{b} The power spectral density (PSD; arbitrary normalization) of the hard X-ray continuum, obtained by Fourier analysis of the time series of the total mass of gas emitting hard X-rays (Appendix \ref{ap:f}). Beyond $150\,\Omega^{-1}(100\,R_g)$, the hard X-ray core precesses at an average angular speed of $0.04\,\Omega(100\,R_g)$ with random fluctuations, producing a characteristic random-walk PSD, $\propto f^{-2}$. Sudden changes in the phase angle in panel \textbf{a} after $250\,\Omega^{-1}(100\,R_g)$ are due to eccentricity variations in the flow embedding the core.\label{fig:varx}}
\end{figure*}

\begin{figure*}
  \centering
  \includegraphics[width=0.8\textwidth]{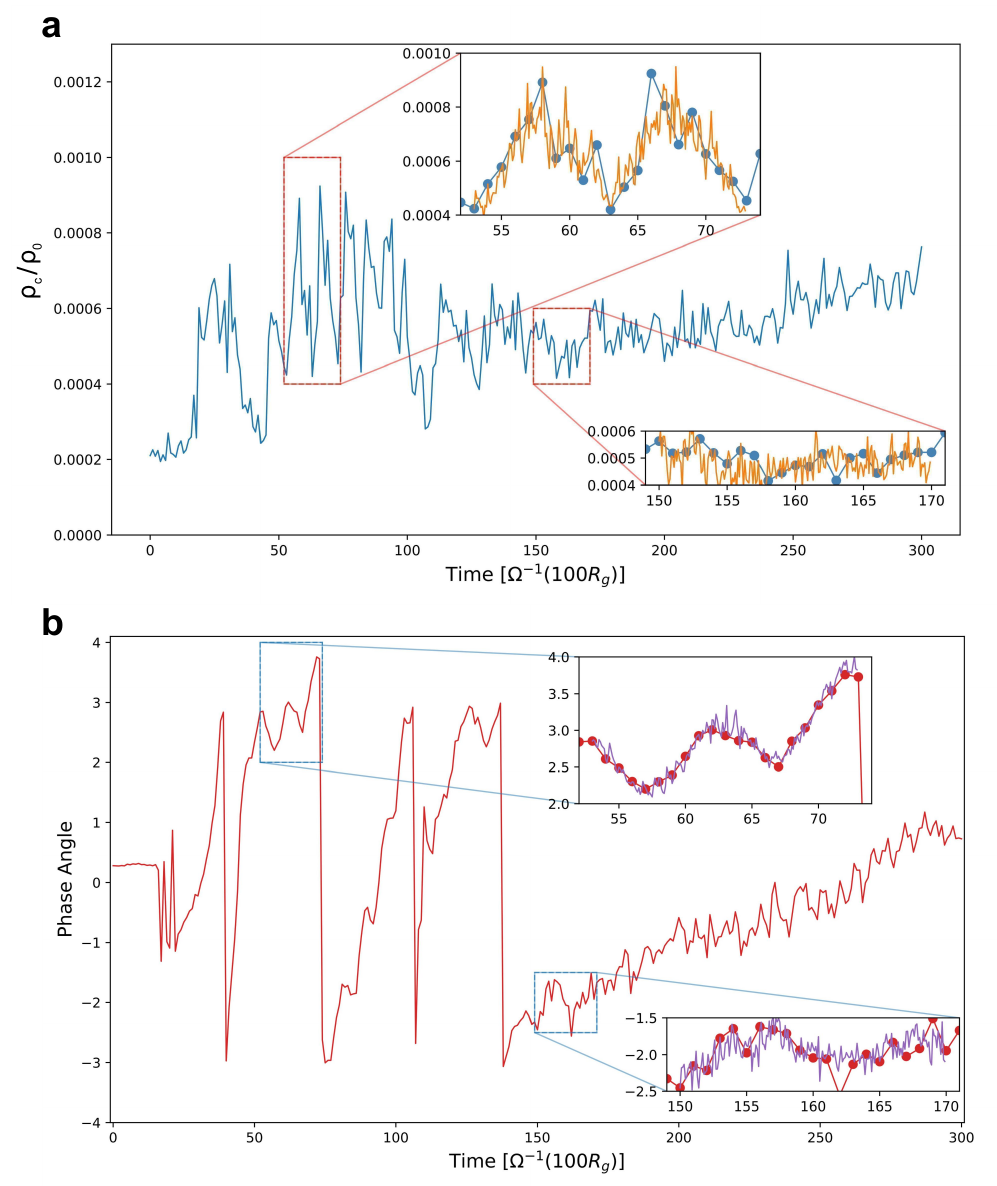}
  \caption{\textbf{Evolution of the X-ray core in A2E05}. Panels \textbf{a} and \textbf{b} show the evolution of the average core density, $\rho_c$, and the core position angle, as in Fig.~\ref{fig:varx}. The inset panels include data from ten-times-higher-cadence reruns of A2E05: A2E05-t1 covering $t=53$--73 and A2E05-t2 covering $t=150$--170 (Table~\ref{tab:1}). A2E05-t2 shows random fluctuations with amplitudes comparable to the low-cadence data, whereas A2E05-t1 confirms coherent bursts in the X-ray region.\label{fig:qpe}}
\end{figure*}
\begin{figure*}
  \centering
  \includegraphics[width=\textwidth]{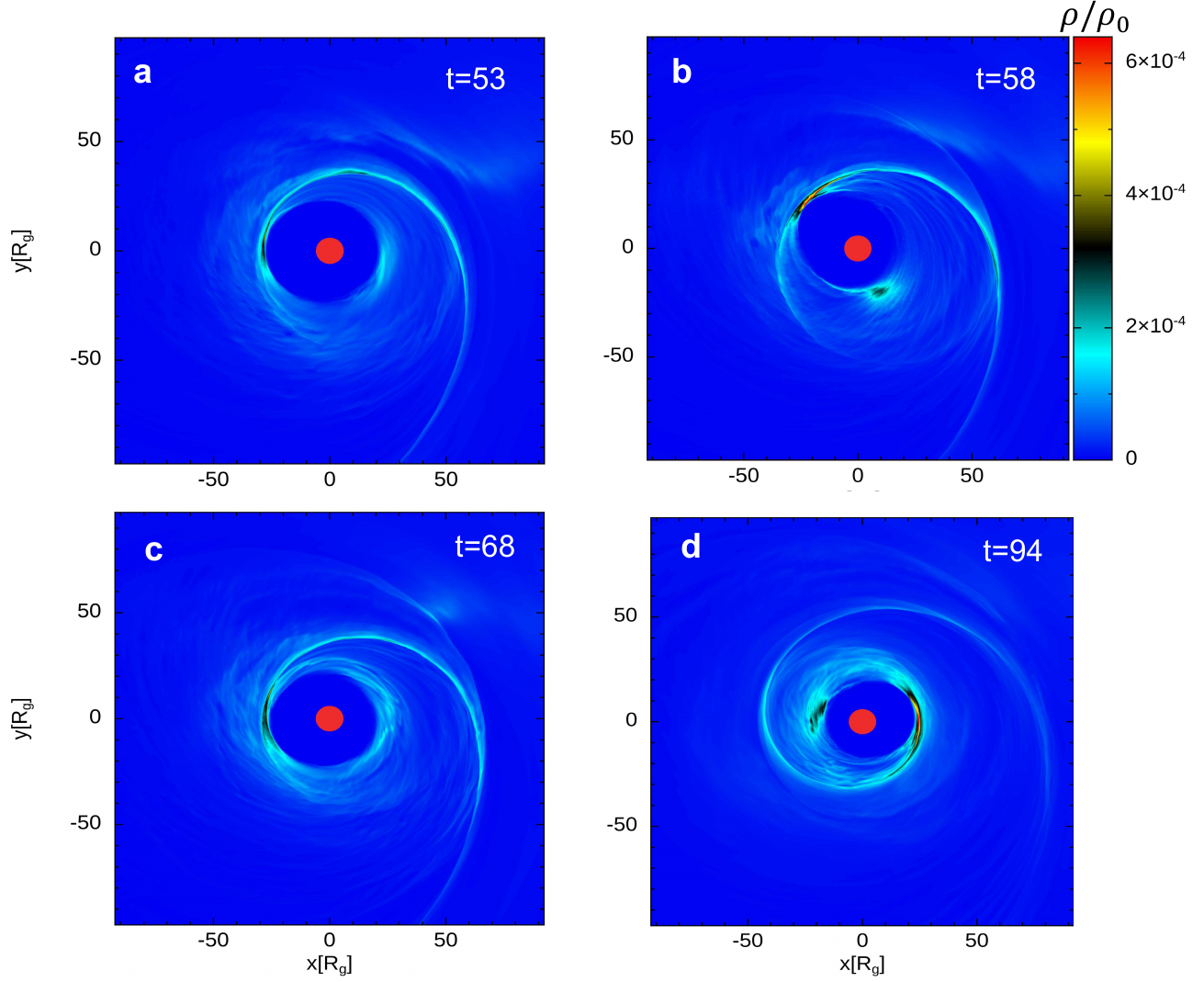}
  \caption{\textbf{Quasi-periodic eruptions}. Panels \textbf{a}--\textbf{d} show the strengthening and weakening of the X-ray core in A2E05 during $t=53$--94$\,\Omega^{-1}(100\,R_g)$. This behavior is associated with the precession of the newly established eccentric flow that embeds the X-ray core.\label{fig:qpe2}}
\end{figure*}
\subsection{X-ray variability}
 As noted above, soft X-rays are typically produced near the periapsis of orbit O1 (Fig.~\ref{fig:rorbits}), whereas the hard X-ray continuum is emitted by a compact core region at $\sim 20\,R_g$ (Fig.~\ref{fig:core}). X-rays can vary on much shorter timescales than the optical lines because GR precession is faster at small radii.
 
 For quantitative analysis, we define the X-ray core as the highest-density region (0.8--1 times the maximum density) and denote its average density by $\rho_c$. We further define the 2--10~keV hard X-ray continuum region as gas with $\rho>\rho_c/8$ and the soft X-ray region as gas with $\rho_c/64<\rho<\rho_c/32$, with the appropriate temperature contrast in our polytropic-gas simulations (Appendix~\ref{ap:f}). This echoes our idealized AGN model in Fig.~\ref{fig:rorbits}, where the most eccentric flow (an O1-like orbit) produces soft X-rays near periapsis (Fig.~\ref{fig:hardsoft}). Variability in the soft X-ray region can propagate inward, driving variations in the hard X-ray continuum. As a result, the low-frequency hard X-ray variability lags behind that of the soft X-ray \cite{kara2025supermassiveaa} (Fig.~\ref{fig:hardsoft}).

Firstly, indirect variability may occur due to variable degrees of obscuration of the hard X-ray core. In the quasi steady state, the core undergoes coherent precession along the most eccentric flow (the O1-orbit analogue), superposed with random drifts in its position angle (Figs.~\ref{fig:varx}a and \ref{fig:qpe}b). These changes can produce obscuration variations down to hour timescales when the line of sight grazes the vertically expanded, low-density broad-line region (the scale height can reach $\sim 10\,R_g$ along orbit O1 when critically twisted; Fig.~\ref{fig:vare}d). This provides a natural explanation for the \emph{cometary shape} and lower-than-expected density of the BLRs in NGC~1365 \cite{maiolino2010comets}. On multi-year timescales, precession of the inner broad-line region (see above) can lead to significant variations in the obscuring column density \cite{zaino2020probing,risaliti2002ubiquitous}.

The core density and X-ray emission are intrinsically variable as well (Fig.~\ref{fig:varx}). During the establishment of the X-ray-emitting region, large-amplitude bursts in peak density/temperature occur due to wobbling of the inner eccentric flow (Figs.~\ref{fig:qpe} and \ref{fig:qpe2}), distinct from the later-stage small, random fluctuations (Fig.~\ref{fig:qpe}). In our compact-disk model A2E05, bursts recur about every $10\,\Omega^{-1}(100\,R_g)$, corresponding to $\sim 50\,\mathrm{ks}$ for a $10^6\,M_\odot$ SMBH. If the feeding flow is compact and thus has a low AMD budget (e.g., a post-TDE eccentric disk), the most compressed region (Fig.~\ref{fig:qpe2}) may still not become hot enough to emit hard X-rays; in that case, the bursts may explain soft X-ray eruptions in newly established accretion flow \cite{miniutti2019nine,hernandez2025discovery}. Further work is needed to determine how the burst pattern and number depend on the eccentric-disk properties.

When the hard X-ray core is fully established, the hard X-ray power spectral density (PSD) scales as $f^{-2}$ (Fig.~\ref{fig:varx}b), as observed in unobscured AGNs \cite{gonzalez2012x}. This naturally explains why more extreme variability is observed around low-mass SMBHs in limited time windows \cite{kara2025supermassiveaa}. X-ray PSDs are also useful for black hole mass measurements because characteristic turnover frequencies scale with black hole mass \cite{mchardy2006active}. In contrast to the well-studied low-frequency turnover \cite{mchardy2006active,gonzalez2012x}, the $f^{-2}$ PSD appears to transition to Poisson noise at a frequency $\nu_t\simeq 1000\,(M_\odot/M)\,\mathrm{Hz}$ for black holes ranging from $10^7\,M_\odot$ \cite[see][Fig.~5]{kara2025supermassiveaa}\footnote{This high-frequency transition may be contaminated by detection noise due to its weak power, but it can be characterized with dedicated observations for nearby bright AGNs.} down to stellar mass~\cite{McClintock2006}. This frequency matches the orbital frequency at $10\,R_g$, where the eccentric flow eventually circularizes (Fig.~\ref{fig:eprofile}). We speculate that eccentric accretion flows ($e\sim 0.5$, excited by companions) in X-ray binaries \cite{Lubow1991,Kley2008simulations} evolve similarly to those in AGNs. Further calibration of this high-frequency PSD turnover could provide a new black hole mass estimator---or rule out this interpretation.

\section{Conclusion and perspectives}
Our results can be summarized as follows:
\begin{enumerate}
\item A moderately eccentric disk with $e\sim 0.5$, as commonly seen in AGN-formation and TDE simulations, naturally develops an ``eccentricity cascade'' and reaches high eccentricity ($>0.8$) at its inner edge, producing extreme vertical compression and a non-axisymmetric temperature structure.
\item The resulting temperature asymmetry provides a physical origin for non-axisymmetric dust sublimation fronts, multiple BLR emitting zones, and the observed UV and soft X-ray excess in AGNs.
\item Differential GR precession drives orbital collisions in the innermost, highly eccentric flow, boosting the gas temperature enough to produce hard X-ray emission within $\sim 20\,R_g$. The hard X-rays exhibit red-noise-like variability that lags the soft X-rays.
\item Precession of the inner eccentric flow on year timescales can trigger collisions with the BLR and produce changing-look events. Precession of the X-ray-emitting core can also yield quasi-periodic eruptions and short-timescale obscuration variability.
\end{enumerate}

This framework also suggests a route toward ``AGN anatomy'' using broad emission lines. With a single well-resolved broad line, one can measure three reverberation lags for the BLR zones \cite{du2016supermassive} and their line-of-sight velocities via profile decomposition \cite{strateva2003double}. These observables provide six constraints that can be used to solve for $(M,a,e,\theta_1,n_x,n_y)$, where $(n_x,n_y,\sqrt{1-n_x^2-n_y^2})$ is the unit vector toward the observer. Once $e$ and $\theta_1$ are inferred, the azimuthal temperature structure follows (Fig.~\ref{fig:rorbits}b). With multiple broad lines, it should be possible to reconstruct a complete map of the AGN structure, although detailed dynamical models may be required for strongly twisted disks. Constraining the emitting areas could, in turn, enable absolute-magnitude estimates of AGNs, turning them into cosmological standard candles.

\clearpage

\begin{acknowledgments}
H.D. is in debt to Gordon Ogilvie for discussions on various aspects of eccentric disk theory. H.D. is grateful to Philip Hopkins, James Stone, Doug Lin, Lucio Mayer, Yaping Li, Shuangliang Li for stimulating discussions on AGN formation and theory, Steve Lubow,  Clement Bonnerot for discussions on eccentric accretion disks around black holes, Hengxiao Guo, Qian Yang, Zhenya Zheng, Pu Du, Haicheng Feng, Wenda Zhang, Zhen Yan, Stefanie Komossa, and Robert Antonucci for stimulating discussions on AGN observations and to Niel Brandt for his wonderful online lectures on AGN observations. Suggestions from an anonymous referee improved the presentation of the paper. H.D. acknowledge support from a talent program of the Chinese Academy of Sciences. We also thank the Chinese Super Computing Centre (Jinan) for their continued support as users of the ShanHe supercomputer, on which all the simulations were performed.
\end{acknowledgments}

\appendix

 \section{Essence of the eccentric disk theory\label{ap:a}}
\subsection{Nonaxisymmetric features}\label{ap:a1}
Accretion disks dominated by elliptical Keplerian motion \cite{ogilvie2001non,ogilvie2014local,ogilvie2019hamiltonian} can be characterized by the eccentricity profile $e(a)$ and the longitude of periapsis $\omega(a)$, which varies continuously with the semimajor axis $a$. The key difference from accretion disks with circular streamlines is that the surface density changes azimuthally as $\Sigma=\Sigma^\circ/j$, where $\Sigma^\circ$ is the surface density of a reference circular disk with the same one-dimensional mass distribution $M_a=2\pi a\Sigma^\circ$, and
\begin{equation}
    j=\frac{1-e(e+ae_a)}{\sqrt{1-e^2}}(1-q\cos\theta)
\end{equation}
is a measure of the surface-area change due to the deformation~\cite{ogilvie2019hamiltonian}. Here $\theta=E-\alpha$, with $E$ being the eccentric anomaly, and $q$ and $\alpha$ satisfying
\begin{equation}
    q\cos\alpha = \frac{ae_a}{1-e(e+ae_a)}, \quad q\sin\alpha= \frac{\sqrt{1-e^2}\,ae\omega_a}{1-e(e+ae_a)}.
\end{equation}
Here $ae_a=a\dd e/\dd a$ and $a\omega_a=a\dd\omega/\dd a$ are the dimensionless gradients of the eccentricity and longitude of periapsis, and $q$ is a measure of the degree of nonlinearity. To avoid orbital crossing, it must satisfy $q<1$~\cite{ogilvie2019hamiltonian}:
\begin{equation}
    1-q^2=\frac{(1-e^2)[1-(e+ae_a)^2-(ae\omega_a)^2]}{[1-e(e+ae_a)]^2}. \label{eq:q}
\end{equation}
Highly eccentric disks are generally prone to orbital collisions, exacerbated by a positive eccentricity gradient. This explains the so-called runaway circularization seen in tidal disruption event simulations, when the inner flow of moderate eccentricity collides with the outer highly eccentric flow \cite{steinberg2024stream}. Orbital crossing also occurs in the inner highly eccentric flow of A2E05-SF (Fig. \ref{fig:eprofile-sf}), leading to shocks and irreversible evolution of the eccentricity.

In addition to horizontal compression, changes in the distance to the central object lead to variations in the vertical component of gravity, forcing the disk to oscillate vertically \cite{ogilvie2001non,ogilvie2014local}. The disk scale height---the standard deviation of the vertical coordinate weighted by density \cite{ogilvie2019hamiltonian}---$H=hH^\circ$ varies by a factor $h$ relative to the reference circular disk, satisfying
\begin{equation}
   (1-e\cos E) \frac{\dd^2 h}{\dd E^2}- e\sin E\frac{\dd h}{\dd E} + h = \frac{(1-e\cos E)^3}{j^{\gamma-1}h^\gamma},
\label{eq:h}
\end{equation}
where $\gamma$ is the adiabatic index. Here and in the following simulations, we adopt a polytropic equation of state with $\gamma=5/3$, since we focus on optically thick flow in the broad-line region, where the hydrogen number density is $\sim 10^{12}\,\mathrm{cm}^{-3}$ \cite{wu2025understanding}. We also adopt $\gamma=4/3$ to mimic the radiation-dominated regime \cite{zanazzi2020eccentric}, which shows no qualitative difference in Fig. \ref{fig:vare}.

We are interested in stationary solutions on the orbital time-scale, which delineates the vertical disk structures well even in evolving eccentric disks (Fig. \ref{fig:H}). Equation \ref{eq:h} is solved via a shooting method by varying $h,\dd h/ \dd E$ to search for $2\pi$ periodic solutions. When there are no twists \cite{zanazzi2020eccentric} ($\omega_a =0$), it is sufficient to find solutions by varying $h$ and enforcing reflection symmetry to the major axis ($\dd h/ \dd E=0$ at $E=0,\pi$). Extreme vertical compression happens near the periapsis \cite{ogilvie2014local,zanazzi2020eccentric} for $e>0.5$  while the maximum of $h$ can reach about 10 (Fig. \ref{fig:vare}d) in highly nonlinear disks with twists ($q \sim 1, \omega_a \neq 0)$. Similar extreme vertical compressions in tidal disruption events (with more complex radial entropy/density profiles) are referred to as \emph{nozzle shocks} \cite{ryu2021impact,bonnerot2022nozzle,shiokawa2015general}. 

\subsection{Long term evolution\label{ap:a2}}
In inviscid flows, the evolution of the eccentricity vector follows~\cite{ogilvie2019hamiltonian,zanazzi2020eccentric}:
\begin{eqnarray}
 M_a\frac{\partial e}{\partial t }&= \frac{\sqrt{1-e^2}}{na^2e}\left[\frac{\partial H_a}{\partial \omega} - \frac{\partial}{\partial a}\left(\frac{\partial H_a}{\partial \omega_a} \right)\right],\label{eq:ecc}\\
    M_a\frac{\partial \omega}{\partial t }&=-\frac{\sqrt{1-e^2}}{na^2e}\left[\frac{\partial H_a}{\partial e} - \frac{\partial}{\partial a}\left(\frac{\partial H_a}{\partial e_a} \right)\right] +M_a   \dot{\omega}_{\rm GR},\label{eq:omega}
\end{eqnarray}
where $n=\sqrt{GM/a^3}$ is the mean motion (we frequently denote it as $\Omega$ in plots, as in circular disks), and $H_a=2\pi a P^\circ F^{\mathrm{(3D)}}$ is the Hamiltonian density, with $P^{\circ}$ the vertically integrated pressure for the reference disk and $F^{\mathrm{(3D)}}$ a geometric factor reflecting the orbit-averaged internal-energy change due to vertical compression \cite{ogilvie2019hamiltonian}. These equations highlight the need to resolve the vertical structure in hydrodynamic simulations. 

When GR precession is negligible, the precession timescale is determined by the gas pressure (see Eq.~\ref{eq:omega}) and is a factor of $(a/H^\circ)^2$ longer than the local dynamical timescale. Still, the precession timescale is a factor of $1/\alpha$ shorter than the viscous timescale \cite{shakura1973black}. We note that one-dimensional evolution equations for eccentric disks, including viscous/turbulent transport, are available in Refs.~\cite{ogilvie2001non,ogilvie2014local}; these require evaluating more complex stress tensors than in traditional circular disks \cite{shakura1973black}. To avoid distraction, we defer detailed analysis of turbulent transport and spectral energy distribution (SED) fitting to another paper, due to its subdominant role in angular momentum transport compared to orbital collisions (Fig. \ref{fig:amd}).

A class of uniformly precessing, untwisted eccentric disks can be found, in which Eqs.~\ref{eq:ecc} and \ref{eq:omega} reduce to an eigenvalue problem for the eccentricity profile satisfying
\begin{equation}
     M_a\omega_0= - \frac{\sqrt{1-e^2}}{na^2e}\left[\frac{\partial H_a}{\partial e} - \frac{\partial}{\partial a}\left(\frac{\partial H_a}{\partial e_a} \right)\right],
\end{equation}
with eigenvalue $\omega_0$ equal to the disk precession frequency \cite{ogilvie2019hamiltonian}. Differentially precessing eccentric disks develop increasing twists with time and become vulnerable to orbital collisions (large $q$; see Eq.~\ref{eq:q}) and dissipation, so they likely evolve toward a uniformly precessing state. However, they may experience a long period of eccentricity oscillation before eventually reaching a uniformly precessing state (Fig. \ref{fig:eprofile-sf}).

Here we use uniformly precessing disks as tests (see Appendix \ref{ap:b2}) for our direct hydrodynamics simulations to find the required resolution for properly capturing disk precession and subtle turbulence due to the parametric instability in eccentric disks~\cite{ogilvie2014local}. We precomputed dense tables for the geometric factor $F^{\mathrm{(3D)}}$ with different $e$ and $e_a$, and found the precession frequency for our specific disk model via shooting methods, forcing torque-free boundary conditions \cite{ogilvie2019hamiltonian,zanazzi2020eccentric}. The Python scripts solving for the vertical scale height, the geometric factor, and the uniformly precessing disks are available at an online repository.

\section{Hydrodynamic simulations\label{ap:b}}
Early two-dimensional simulations that neglected gas pressure suggested that eccentric disks can be long-lived~\cite{Syer1992}. However, three-dimensional disks have a different Hamiltonian, and therefore evolution (Appendix~\ref{ap:a2}) due to a vertical flow modulation ~\cite{ogilvie2019hamiltonian}. The extreme vertical compression (by a factor of $>100$, even for $e\sim 0.6$; Fig.~\ref{fig:H}) can easily fall below the resolution scale of an Eulerian grid code~\cite{chan2024three}, posing a challenge for fully three-dimensional hydrodynamic simulations. We therefore adopt the meshless finite mass (MFM) scheme in the GIZMO code~\cite{hopkins2015new}, a Godunov-type Lagrangian method. MFM performs an effective volume partition given a distribution of fluid elements (particles) and updates fluid variables using fluxes obtained from Riemann solvers. It can naturally follow complex flows in distorted accretion disks and resolve subtle parametric instabilities thanks to its hybrid nature and the absence of artificial viscosity~\cite{deng2022non}.

\subsection{Initial conditions}
We use code units of 1 au, 1 million solar masses, and $G=1$. We note that only the relative strength of gas pressure to orbital energy (i.e., disk thickness) matters for all the following simulations, even with GR precession (see Appendix~\ref{ap:a2}); the results can be rescaled and applied to more extended, lower-density gas.

To save computational resources, we construct compact reference circular-disk models covering $[a_0,2a_0]$ or $[a_0,3a_0]$, where $a_0$ is the code length unit. The disk has a mass distribution $M_a\propto\tanh\!\big(\frac{a-a_0}{0.1a_0}\big)\tanh\!\big(\frac{2a_0-a}{0.15a_0}\big)$ and a total mass of $10^{-5}$ in code units. We adopt a polytropic equation of state (EOS),
\begin{equation}
    P=A_0\left( \frac{\rho}{\rho_0}\right )^{5/3},
\end{equation}
where $A_0$ is a constant and $\rho_0$ is the code density unit. By that, we assume no entropy gradient and instantaneous loss of shock-generated entropy. We choose $A_0=2.11$ such that the disk scale height is $\sim 0.033a$; the disk is intentionally thick to allow good vertical resolution. With a polytropic EOS, the disk has a finite vertical extent (Figs.~\ref{fig:H} and~\ref{fig:pi}), transitioning to vacuum at high altitudes. No explicit boundary condition is imposed, since our Lagrangian code can follow arbitrary deformation of the disk surface.

We generate a particle representation of the density profile via rejection sampling and relax the disk by damping radial and vertical velocity fluctuations due to numerical noise, reaching an equilibrium state~\cite{deng2022non}. We then deform the reference circular disk into the desired eccentric-disk models, given specific eccentricity profiles (e.g., the uniformly precessing profile in Appendix~\ref{ap:a}), using the following steps: (1) divide the circular disk into 200 radial bins; (2) rescale the vertical coordinate according to the solution of Eq.~\ref{eq:h} with the desired eccentricity in each bin; (3) add the vertical oscillation velocity; (4) deform the circular bins into elliptical orbits; and (5) adjust the orbital velocity to the desired elliptical Keplerian motion. Similar approaches are used in the construction of nonlinear, steady warped accretion disks~\cite{deng2022non}. A Python script for this transformation is available at an online repository.

\subsection{Numerical convergence}\label{ap:b2}

We employ 5 million or 42 million particles in the pre-deformed circular-disk model (we often split particles in the low-resolution model to reach higher resolution~\cite{deng2022non}), which gives a midplane resolution of $H^\circ/5$ and $H^\circ/10$, respectively. We test numerical convergence against semi-analytical theory.

First, we aim to capture the disk precession rate accurately, which requires resolving the disk vertical structure (Appendix~\ref{ap:a2}). In the low-eccentricity uniformly precessing disk (Fig.~\ref{fig:upress}), 5 million particles are sufficient to reproduce the coherent precession rate to within $1\%$. In the twisted, high-eccentricity disk, our 42 million particle simulation is sufficient to resolve the extreme vertical compression (Fig.~\ref{fig:H}) and hence the precession.

Capturing the more delicate parametric instability---due to resonance between pairs of inertial waves and the background eccentric flow~\cite{barker2014hydrodynamic}---requires even higher resolution. A similar parametric instability in a warped disk required a midplane resolution of $H/8$~\cite{deng2021parametric}. Consistent with this, we find that the parametric instability occurs only in our high-resolution (42 million particle) simulation of the uniformly precessing model (Fig.~\ref{fig:upress}).

As a result, we adopt 42 million particles as our fiducial resolution. A more extended disk of similar thickness but twice as massive, covering $[a_0,3a_0]$ (with the same taper functions at the edges), is constructed with 51 million particles to allow a larger dynamical range (see Fig.~\ref{fig:r3}). These simulations are among the best-resolved accretion-disk simulations using the MFM method so far. Nevertheless, the parametric instability in highly eccentric flows may still not be fully captured.

\section{The central potential and inclusion of GR precession\label{ap:c}}

We present two classes of simulations, as summarized in Table \ref{tab:1}: scale-free simulations intended to reveal general eccentric-disk evolution, and simulations with GR precession to investigate black hole accretion.

In the scale-free simulations, we include only the Keplerian potential of a supermassive black hole of one million solar masses (the code mass unit). These simulations reveal the evolution of eccentric disks far from the black hole, where GR effects can be ignored. The results can be rescaled (regardless of the code units used here) and applied to a wide range of radii in AGNs.

The high-eccentricity scale-free simulations, A2E05-SF/A3E05-SF, form highly eccentric flow in the inner disk, leading to orbital collisions and rapid accretion. When the flow reaches the vicinity of the black hole ($\sim 100R_g$, where $R_g=GM/c^2$ is the gravitational radius), GR effects become important~\cite{tejeda2013accurate}. A good approximation for GR precession is crucial because it largely determines the precession of eccentric flows (affected by gas pressure) and drives disk variability. We adopt the post-Newtonian potential proposed in Ref.~\cite{tejeda2013accurate}, which exactly predicts the periapsis advance of elliptical orbits. The correction terms, whose strength is characterized by $R_g$, are added to the central Keplerian potential. We confirm with numerical tests that our implementation reproduces the periapsis advance of test particles.

To mimic and accelerate the inspiral of material (Fig. \ref{fig:eprofile-sf}cd), we increase $R_g$ from 0 to $0.01a_0$ over $10\pi\,\Omega^{-1}(a_0)$ according to
\begin{equation}
    R_g=0.01a_0\sin^2\left(\frac{t}{20}\right).
\end{equation}

\section{Angular momentum deficit and orbital energy\label{ap:d}}

The angular momentum deficit (AMD) of a fluid element of mass $m_k$ on an eccentric orbit with semimajor axis $a_k$, eccentricity $e_k$, and inclination $i_k$ is
\begin{equation}
    m_k\sqrt{GMa_k}\left(1-\sqrt{1-e_k^2}\cos i_k\right).
\end{equation}
The total AMD of the disk is the sum of these quantities and is conserved (Fig.~\ref{fig:amd}) in the absence of shocks and turbulence \cite{deng2021parametric}. We track the AMD to show how the limited initial orbital excitation is damped in the absence of external AMD supply.

For the gravitational potential we adopt, the orbital energy \cite{tejeda2013accurate} of a fluid element is
\begin{equation}
    \frac{1}{2}\left(\frac{r_k}{r_k-2R_g}\right)^2 m_k v_{r,k}^2+\frac{1}{2}\left(\frac{r_k}{r_k-2R_g}\right) m_k v_{t,k}^2-\frac{GMm_k}{r_k},
\end{equation}
where $r_k$ is the distance to the central object and $v_{r,k}$ and $v_{t,k}$ are the radial and tangential velocities of the fluid element. The gas internal energy is negligible compared to the orbital energy, and the released orbital energy is efficiently lost at shock fronts under our polytropic equation-of-state assumption.

\section{Reynolds stress and effective viscous parameter\label{ap:g}}

Turbulent velocity fluctuations produce effective stresses that transport angular momentum \cite{balbus1998instability}. In eccentric disks, the specific angular momentum is not constant at a given cylindrical radius $r$, so turbulent transport is better characterized in the $(\lambda,\phi)$ plane \cite{ogilvie2014local}, where
\begin{equation}
   r=R(\lambda,\phi)=\frac{\lambda}{1+e(\lambda)\cos\!\left[\phi-\omega(\lambda)\right]}.
\end{equation}
Here $\lambda=a(1-e^2)$ is a quasi-radial coordinate that labels orbits of constant specific angular momentum; $h=\sqrt{GM\lambda}$. The contravariant velocity components $(v^\lambda,v^\phi)$ are related to $(v^r,v^\phi)$ via the Jacobian $J_1=\frac{\partial(\lambda,\phi)}{\partial(r,\phi)}$, giving
\begin{equation}
 v^\lambda=\frac{v_r}{R_\lambda}-\frac{R_\phi}{R_\lambda}v^\phi.
\end{equation}
Here $v^\phi$ is the angular velocity, and the partial derivatives $R_\lambda$ and $R_\phi$ are given in \cite{ogilvie2014local}. The contravariant stress tensor is related to that in polar coordinates by $\bar{T}=J_1 T J_1^{T}$, and
\begin{equation}
    T^{\lambda\phi}=\frac{1}{R_\lambda}T^{r\phi}-\frac{R_\phi}{R_\lambda}T^{\phi\phi},
\end{equation}
with
\begin{equation}
    T^{r\phi}=\rho\,\delta v^r\,\delta v^\phi,\quad T^{\phi\phi}=\rho\,\delta v^{\phi}\,\delta v^{\phi}.
\end{equation}
The physical stress tensor (with proper units) can be recovered using the metric tensor as
\begin{equation}
    T_{\lambda\phi}=\sqrt{g_{\lambda\lambda}g_{\phi\phi}}\,T^{\lambda\phi}=\sqrt{R^2+R_\phi^2}\,\left(T^{r\phi}-R_\phi T^{\phi\phi}\right).
\end{equation}
We compute the volume-averaged $T_{\lambda\phi}$ over $1.2a_0<\lambda<1.3a_0$ after saturation of the parametric instability (Fig.~\ref{fig:pi}), and normalize it by the volume-averaged pressure to obtain an effective viscous $\alpha$ parameter~\cite{shakura1973black}. We find that $\alpha$ is highly time-dependent in the idealised uniformly precessing model A2UP-SF (HR), with mean $<10^{-3}$ and standard deviation $4\times 10^{-3}$. However, it remains unclear whether the $\alpha$-viscosity prescription applies to eccentric accretion disks~\cite{ogilvie2001non,Lyubarskij}. In fact, we calculated a similar $\alpha$ in A2E05-SF, but the angular momentum transport is evidently faster than that in A2UP-SF.

\section{On the role of radiation pressure and magnetic fields}\label{ap:e}

The temperature is proportional to $(jh)^{1-\gamma}$, and the ratio of radiation pressure to gas pressure is proportional to $(jh)^{4-3\gamma}$; both therefore vary azimuthally, in contrast to standard circular-disk theory. Gas pressure dominates over radiation pressure along the reference orbit O2 in Fig.~\ref{fig:rorbits}. By contrast, radiation pressure can dominate for material interior to orbit O1. However, reducing the minimum temperature by a factor of 4 (still sufficient to produce a soft X-ray excess near periapsis, allowing higher eccentricity or twist; Fig.~\ref{fig:vare}) while increasing the density by a factor of 10 in this inner region---plausible if the quoted BLR densities \cite{wu2025understanding} reflect conditions near the disk surface---can make radiation pressure subdominant.

The current model is not yet complete: our treatment of GR is approximate, and we neglect radiation pressure and magnetic fields, which are likely required for accretion within $10\,R_g$ (Fig.~\ref{fig:hardsoft}) and for jet formation. Nevertheless, the gravitational energy is much larger than both the radiation and gas pressure and controls the orbital motion. Our results therefore provide useful information on how material is dynamically fed into the $<100\,R_g$ region (along eccentric orbits featuring extreme compression), which is typically treated as an isolated torus in radiation magnetohydrodynamic (MHD) simulations \cite{jiang2014global}. The universal soft X-ray excess (Figs.~\ref{fig:rorbits} and \ref{fig:vare}) and the characteristic X-ray power spectrum (Fig.~\ref{fig:varx}) are reasonably explained, suggesting that the orbital dynamics of the eccentric flow largely drive the variability.

On the other hand, to the best of our knowledge, the only dedicated MHD simulations of eccentric disks \cite{chan2024three} find subsonic magnetorotational-instability (MRI) turbulence that is weaker than the parametric-instability turbulence, as measured by the Mach number of the velocity fluctuations (Fig.~\ref{fig:pi}). This is consistent with the subdominant role of MRI turbulence in warped disks \cite{fairbairn2025interplay}. These simulations also highlight enhanced magnetic reconnection in vertically compressed regions, though it is poorly resolved numerically, which weakens the magnetic field. We note that the magnetic stress in Ref.~\cite{chan2024three} is not calculated for flow of constant angular momentum~\cite{ogilvie2014local} and thus not directly comparable to the Reynolds stress in Appendix \ref{ap:g}.

Further GRMHD simulations---ideally with aggressive adaptive mesh refinement at periapsis (Fig.~\ref{fig:H})---are warranted to assess the role of magnetic fields in eccentric flows near black holes; radiation should be included when possible.

\section{Analysis of the X-ray emission regions\label{ap:f}}

To better quantify variability in eccentric disks around SMBHs, we carry out time-series analyses of the X-ray-emitting regions. Only the relative density matters in our simulations, and denser regions are hotter under our polytropic equation of state ($T\propto \rho^{2/3}$). The core region (with density 0.8--1 times the peak density) is typically compact, so its position and precession can be readily determined (Figs.~\ref{fig:varx} and \ref{fig:qpe}). In rare cases, the core becomes azimuthally extended and embedded in low-eccentricity flow, leading to sudden changes in the inferred core phase angle (Fig.~\ref{fig:varx}a, around $250\,\Omega^{-1}(100\,R_g)$).

The peak temperature reached near the black hole is largely set by the supplied material along the reference orbit O1 (Fig.~\ref{fig:rorbits}), linking optical and X-ray emission \cite{lusso2016tight}. Although the temperature along O1 is affected by irradiation from the central engine \cite{fausnaugh2016space} and by the assumed eccentricity (Fig.~\ref{fig:vare}), observationally there appears to be a nearly universal blackbody soft X-ray excess at $\sim 0.1\,\mathrm{keV}$ \cite{bianchi2009caixa}. We associate this soft X-ray excess with hot gas near the periapsis of O1, and associate the hard X-ray continuum (2--10~keV) with regions of density up to 64 times higher (Fig.~\ref{fig:hardsoft}). The peak gas density varies with time (Fig.~\ref{fig:varx}), as does the core density $\rho_c$ (defined as the mean density of particles with 0.8--1 times the peak density). To remove this uncertainty, we use the time-averaged core density $\langle\rho_c\rangle$ as a reference: the soft X-ray region is defined by $\langle\rho_c\rangle/64<\rho<\langle\rho_c\rangle/32$, and the hard X-ray region by $\rho>\langle\rho_c\rangle/8$.

We compute time series of the total mass within each density-defined region as a proxy for the luminosity. We then Fourier-analyze these time series to obtain the PSD (Fig.~\ref{fig:varx}) and the hard--soft phase lags as a function of frequency (Fig.~\ref{fig:hardsoft}). While the hard X-ray lag can be explained by inward propagation of perturbations \cite{kara2016global}, we lack reprocessing models to explain the leading hard X-ray variations at high frequencies \cite{fabian2009broad}. We note that relativistic effects---such as light bending, beaming, and gravitational redshift---may change the observed X-ray properties; this should be assessed with relativistic ray-tracing studies~\cite{brenneman2006constraining}.

This correspondence assumes that orbit O1 is sufficiently eccentric (Fig.~\ref{fig:vare}) to compress gas at periapsis to temperatures high enough for soft X-ray blackbody emission, which should be common in AGNs (Fig.~\ref{fig:rorbits}). This may not hold for post-TDE disks with limited dynamical range and AMD budget. In that case, the periapsis flow in O1 may be too cool to emit soft X-rays, and only the final GR-precession-driven compression can raise it to soft X-ray-emitting temperatures (Figs.~\ref{fig:qpe} and \ref{fig:qpe2}).

\end{document}